\documentclass[preprintnumbers,prd,twocolumn,showpacs,floatfix,preprintnumbers,superscriptaddress,nofootinbib]{revtex4-2}

\usepackage{graphicx}
\usepackage{epsfig}
\usepackage{bm}
\usepackage{amssymb}
\usepackage{float}
\usepackage{amsmath}
\usepackage{subfigure}
\usepackage{dcolumn}
\allowdisplaybreaks
\usepackage{multirow}
\usepackage[colorlinks]{hyperref}
\usepackage[usenames,dvipsnames]{color}
\hypersetup{
     breaklinks=true,
    pdfstartview={FitH},    
    colorlinks=true,       
    linkcolor=blue,          
    citecolor=red,        
    filecolor=magenta,      
    urlcolor=blue,           
    anchorcolor=green,      
    linktocpage=true
}
\usepackage{orcidlink}

\newcommand{\Mpl}{M_{\textrm{Pl}}}

\newcommand{\nn}{\nonumber}

\def\al{\alpha}

\def\doi{http://doi.org}

\allowdisplaybreaks

\begin{document}

\title{A single field inflationary model consistent with recent observations: reheating and radiative stability}

\author{Md. Wali Hossain\orcidlink{0000-0001-6969-8716}}
\email{mhossain@jmi.ac.in}
\affiliation{Department of Physics, Jamia Millia Islamia, New Delhi, 110025, India}

\begin{abstract}
We study single field inflation with the inverse exponential potential, which has one free parameter besides the energy scale. For a wide range of parameter space, its predictions for the scalar spectral index and the tensor to scalar ratio lie within the $1\sigma$ region of the combined SPA$+$BK$+$DESI data, and the wider family of inverse power law exponential potentials behaves similarly. We compare these predictions with those of the $R^2$ model, monomial potentials, $\alpha$-attractors, non-minimally coupled Higgs like inflation and a radiatively corrected plateau potential. The $R^2$ and monomial predictions lie near or outside the $2\sigma$ boundary of the same data, whereas the non-minimally coupled Higgs like and radiatively corrected plateau families overlap the favoured contours over part of their parameter space. To describe reheating, we add a steep exponential term that is negligible during inflation and, together with the inverse exponential term, creates a minimum about which the inflaton oscillates. The maximum reheating temperatures are $T_{\rm rh}\sim10^{12}$--$10^{13}\,{\rm GeV}$, consistent with the Big Bang Nucleosynthesis bound. Coupling the inflaton to a Dirac fermion through a Yukawa interaction, these temperatures require couplings comparable to the standard model charged lepton Yukawa couplings. Nonperturbative fermion production transfers negligible energy, and perturbative decay completes reheating. The same coupling induces a one loop correction to the inflationary potential, which limits the radiatively safe reheating temperature to $T_{\rm rh}\lesssim10^{12}\,{\rm GeV}$.
\end{abstract}

\maketitle

\flushbottom

\section{Introduction}
\label{sec:intro}

Inflation \cite{Guth:1980zm,Starobinsky:1980te,Linde:1981mu,Linde:1983gd} generically predicts a nearly scale-invariant spectrum of primordial perturbations, characterized by the scalar spectral index $n_{\rm s}$. Measurements by the Planck satellite 2018 release (P) gave $n_{\rm s}=0.9651\pm0.0044$ \cite{Planck:2018jri,Planck:2018vyg}. More recent observations slightly modify this picture. The Atacama Cosmology Telescope sixth data release (ACT DR6) reports $n_{\rm s}=0.9666\pm0.0077$ \cite{AtacamaCosmologyTelescope:2025blo,AtacamaCosmologyTelescope:2025nti}. Combining Planck and ACT (P-ACT) with data from the Dark Energy Spectroscopic Instrument Data Release 1 and 2 (DESI DR1, DR2) \cite{DESI:2024uvr,DESI:2024mwx,DESI:2025zgx} shifts the preferred value toward larger values, yielding $n_{\rm s}=0.9743\pm0.0034$ and $0.9752\pm0.0030$, respectively \cite{AtacamaCosmologyTelescope:2025blo}, nearly $2\sigma$ higher than the original Planck result. Independent measurements from the South Pole Telescope third-generation camera (SPT-3G) give $n_{\rm s}=0.951\pm0.011$ \cite{SPT-3G:2025bzu}, consistent with Planck within uncertainties. The SPA combination of SPT-3G, Planck, and ACT yield $n_{\rm s}=0.9684\pm0.0030$, which increases to $0.9728\pm0.0027$ when DESI DR2 data are included \cite{SPT-3G:2025bzu}. Incorporating data from the BICEP/Keck Array (BK) \cite{BICEP:2021xfz} gives $n_{\rm s}=0.9682\pm0.0032$ for SPA-BK and $0.9728\pm0.0029$ for SPA-BK-DESI2, which also constrains the tensor-to-scalar ratio to $r<0.035$ \cite{Balkenhol:2025wms}.

These updated constraints, particularly those including DESI data, place increasing pressure on standard single-field inflationary scenarios. The $R^2$ Starobinsky model \cite{Starobinsky:1980te,Starobinsky:1983zz}, previously favoured by Planck 2018 \cite{Planck:2018jri}, lies near the boundary of the $2\sigma$ allowed region of $n_{\rm s}$ from P-ACT-BK-DESI1 \cite{AtacamaCosmologyTelescope:2025nti} and SPA-BK-DESI2 \cite{Balkenhol:2025wms} for $N=60$ e-folds and becomes disfavoured at more than $2\sigma$ for smaller $N$. Monomial models $\phi^n$ \cite{Linde:1981mu,Linde:1983gd} accommodate SPA-BK-DESI2 only at the $2\sigma$ level for sufficiently small powers, roughly $n<1/3$. Consequently, a large number of recent works have explored modifications or extensions of standard scenarios to reconcile theory with current observations. These include minimally coupled single-field models with modified potentials \cite{Lynker:2025wyc,Ai:2025qua,Abhisheki:2025kvw}, generalized Starobinsky and plateau models \cite{Drees:2025ngb,Gialamas:2025ofz,Kouniatalis:2025orn,Racioppi:2025igu,Haque:2025uga}, non-minimally coupled or modified-gravity realizations including Higgs-like non-minimal couplings \cite{Aoki:2025wld,Salvio:2025izr,Zharov:2025zjg,Gao:2025onc,Yin:2025rrs,Liu:2025qca,McDonald:2025tfp,Kumar:2025apf,Herrera:2026mhr,Odintsov:2025wai}, $\alpha$-attractor models \cite{Dioguardi:2025vci,Dioguardi:2025mpp,He:2025bli,Han:2025cwk,Wolf:2025ecy,Heidarian:2025drk,Kallosh:2025sji,Singh:2025uyr,Oikonomou:2026vrw,Kallosh:2026tke,Jing:2026ymp}, Warm inflation scenarios \cite{Berera:2025vsu,Chakraborty:2025jof,Das:2025teu,Ahmed:2026msg}, quantum and radiative corrections \cite{Kim:2025dyi,Ahmed:2025rrg,Gialamas:2025kef,Yuennan:2025kde,Bianchi:2025tyl,Modak:2025bjv,Ahmed:2025sfm,Odintsov:2026doe} and other scenarios including chaotic, curvaton, hybrid, supersymmetric, string inspired, and primordial black hole motivated inflationary models \cite{Balkenhol:2025wms,Kallosh:2025rni,Byrnes:2025kit,Rehman:2025fja,Kallosh:2025ijd,Peng:2025bws,Maity:2025czp,Leontaris:2025hly,Ferreira:2025lrd,Chakraborty:2025wqn,Allegrini:2025jha,Aldabergenov:2025kcv,Keus:2025iwa,Wang:2025cpp,Yi:2025dms,Frolovsky:2025iao,Santos:2026ojg,Zhang:2026ivx,Sabogal:2026qvy}. Despite these developments, it remains challenging to construct a minimally coupled single-field inflationary model that is both simple and consistent with the recent SPA-BK-DESI2 constraint on $n_{\rm s}$. In this work, we consider a canonical scalar field with a monotonic inflationary potential whose inflationary dynamics are determined by a single free parameter, apart from the overall energy scale. An additional parameter is introduced only when extending the model to describe the reheating phase.

Motivated by this, we consider an inverse exponential (IExp) potential of the form
\begin{equation}
V(\phi)=V_0 e^{-\alpha M_{\rm Pl}/\phi},
\label{eq:potential_inflation}
\end{equation}
where $\alpha$ is a dimensionless parameter. This potential is assumed to be valid during inflation for $\phi>0$. The IExp potential is a particular subclass of the more general potential \cite{Geng:2015fla}
\begin{equation}
V(\phi)=V_0 e^{-\alpha (\phi/M_{\rm Pl})^n},
\label{eq:potential_inflation_general}
\end{equation}
where $n$ is a constant parameter and the IExp potential corresponds to the case $n=-1$. Potential~\eqref{eq:potential_inflation_general} was introduced as a phenomenological minimal extension of the exponential potential.

In \cite{Geng:2015fla}, the general potential~\eqref{eq:potential_inflation_general} was studied for $n>0$, which represents a class of generalized steep exponential potentials. However, the case $n>0$ cannot produce viable values of the $n_{\rm s}$ consistent with current observations, such as those from SPA$+$BK$+$DESI. Moreover, obtaining sufficiently small values of the $r$ requires extremely small values of the parameter $\alpha$, typically $\alpha \sim 10^{-8}$ \cite{Geng:2015fla}. In this work we show that these issues can be resolved by considering the regime $n<0$. In particular, this allows for a viable single field inflationary scenario consistent with current observational constraints. For simplicity, we focus on the specific case $n=-1$, which corresponds to the IExp~\eqref{eq:potential_inflation} potential and effectively reduces the model to a single parameter potential.

\begin{figure}[ht]
\centering
\includegraphics[scale=0.32]{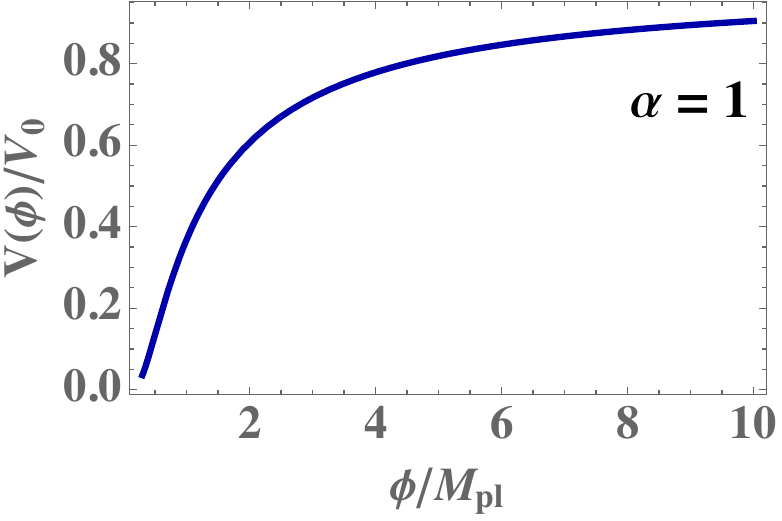}
\includegraphics[scale=0.32]{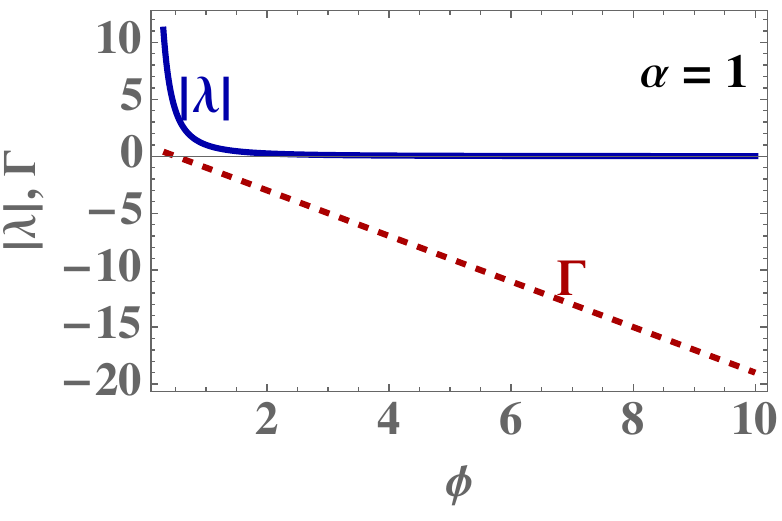}
\caption{Nature of the potential~\eqref{eq:potential_inflation} and its slope ($\lambda$) and curvature ($\Gamma$) for $\phi/M_{\rm Pl} > 0$ with $\alpha = 1$.}
\label{fig:potential_inflation}
\end{figure}

We analyse the slope and the curvature parameters, $\lambda\equiv-M_{\rm Pl}V'/V$ and $\Gamma\equiv VV''/V'^2$, which are related to the slow roll parameters $\epsilon_V$ and $\eta_V$ as
\begin{align}
    \epsilon_V &= \frac{\lambda^2}{2} \, , \qquad \eta_V = \lambda^2\Gamma\, .
    \label{eq:lamGamEpEta}
\end{align}
Observationally viable inflation requires a small $|\lambda|$ together with moderate to large negative $\Gamma$ (see Appendix~\ref{app:lambdaGamma}). $\Gamma$ is the factor converting the square of the slope into the curvature of the potential, and its value controls the relative growth of $\eta_V$ against $\epsilon_V$ as the field evolves.

For $\Gamma = 1$, the potential reduces to a pure exponential, $V \propto e^{-\lambda \phi/M_{\rm Pl}}$, for which $\eta_V = 2\epsilon_V$ identically. In this limit the field asymptotes to the power law inflation attractor without a natural exit. A single exponential sector therefore cannot terminate inflation on its own.

In the strongly concave regime, $\Gamma \ll 0$, Eq.~\eqref{eq:lamGamEpEta} shows that $\eta_V$ can nevertheless remain small and negative, provided $\lambda$ is sufficiently small, as required for a nearly flat slow-roll potential. For representative values of $r$ and $n_{\rm s}$, the corresponding values of $\epsilon_V$, $\eta_V$, $\lambda$, and $\Gamma$ are listed in Table~\ref{tab:lambda-gamma-values} of Appendix~\ref{app:lambdaGamma}. The table shows that the degree of concavity of an inflationary potential is more transparently characterized by $\Gamma$ than by $\eta_V$, since the latter is additionally suppressed by the smallness of $\lambda^2$ and typically takes values much closer to zero. For concave potentials, generally, $\Gamma$ increases as the field rolls down the potential and eventually becomes positive before reheating if the potential has a minimum after the end of inflation. 

For the potential~\eqref{eq:potential_inflation},
\begin{equation}
\lambda = -\alpha\frac{\Mpl^2}{\phi^2}\, , \qquad \Gamma = 1 - \frac{2}{\alpha}\frac{\phi}{\Mpl} \, ,
\end{equation}
which vanishes at $\phi = \alpha\Mpl/2$. This point is an inflection point of the potential, $V''(\alpha/2) = 0$, and separates two regimes: $\Gamma > 0$ ($V'' > 0$, locally convex) for $\phi < \alpha/2$, and $\Gamma < 0$ ($V'' < 0$, locally concave) for $\phi > \alpha/2$. For $\alpha>0$ and sufficiently large $\phi$, the value of the slope ($|\lambda|$) becomes small while $\Gamma$ attains large negative values, yielding a strongly concave potential that naturally supports slow-roll inflation consistent with current observational requirements. The resulting behaviour of the potential and the parameters $\lambda$ and $\Gamma$ is illustrated in Fig.~\ref{fig:potential_inflation}.

The potential~\eqref{eq:potential_inflation} has an essential singularity at
$\phi=0$: the one-sided limits differ ($V\to0$ as $\phi\to0^{+}$, while
$V\to\infty$ as $\phi\to0^{-}$), and the exponent $\alpha M_{\rm Pl}/\phi$
itself is undefined at $\phi=0$, so $V$ cannot be continuously extended to
this point. This is why the potential is restricted to $\phi>0$ throughout,
and why the theory as defined by Eq.~\eqref{eq:potential_inflation} alone
cannot describe the post-inflationary evolution of the field down to
$\phi=0$, modifications are required to extend the dynamics through this
point, as discussed below.

The paper is organized as follows. Sec.~\ref{sec:consistency} compares the IExp predictions with the SPA$+$BK$+$DESI constraints and Sec.~\ref{sec:comparison} compares them with other model classes. Sec.~\ref{sec:reheating} introduces the extension of the potential that creates a minimum and studies the post-inflationary evolution. Sec.~\ref{sec:Trh} derives the reheating temperature, couples the inflaton to a Dirac fermion to examine perturbative decay and early fermion production, and computes the one loop correction that this coupling induces on the inflationary potential. The conclusions are in Sec.~\ref{sec:conc}.

\begin{figure}[t]
\centering
\includegraphics[scale=0.5]{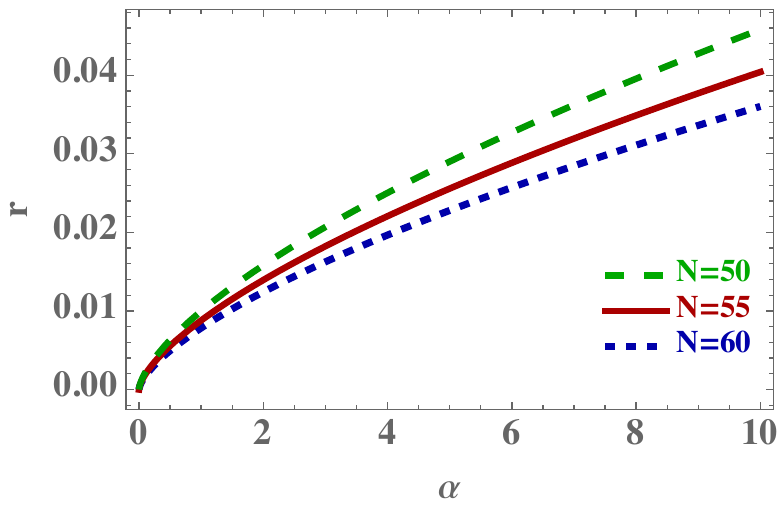}
\includegraphics[scale=0.5]{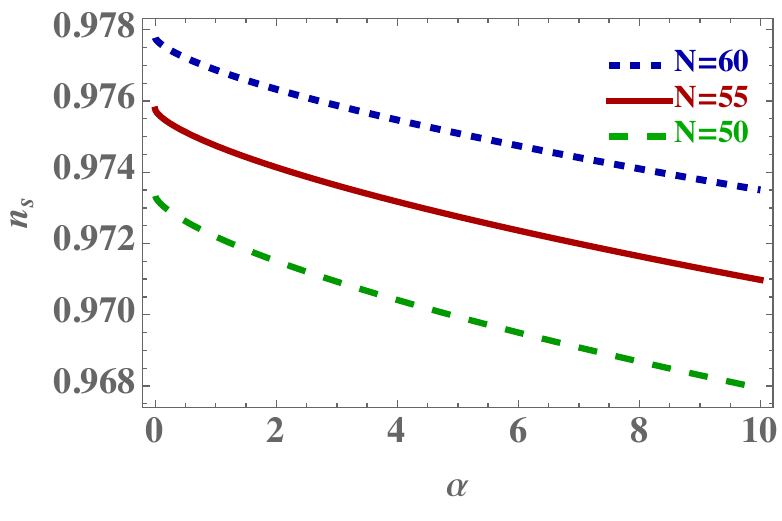}\\
\includegraphics[scale=0.55]{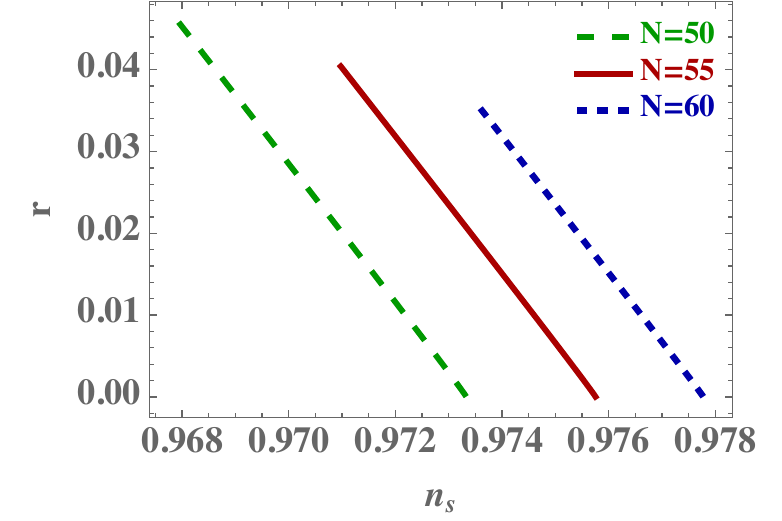}
\includegraphics[scale=0.5]{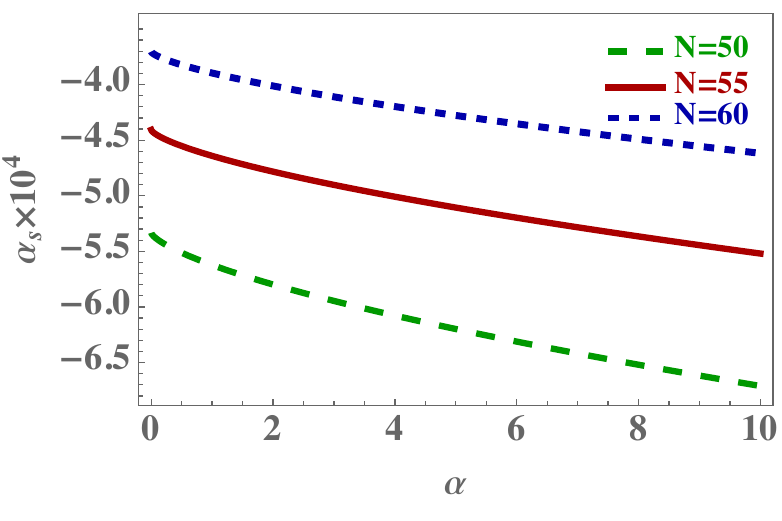}
\caption{Inflation parameters for the potential~\eqref{eq:potential_inflation} for $\phi/\Mpl>0$. For all the plots $0.05 \leq \alpha \leq 10$.}
\label{fig:inflation_parameters}
\end{figure}

\section{Consistency with Observations}
\label{sec:consistency}

The potential slow-roll parameters for the potential~\eqref{eq:potential_inflation} take the explicit form
\begin{align}
    \epsilon_V(\phi) &
    = \frac{\al^2 M_{\rm Pl}^4}{2\phi^4},~~~ 
    \eta_V(\phi) = \frac{\al \Mpl^3}{\phi^3}\left(\frac{\al\Mpl}{\phi}-2\right), 
    \label{eq:ep_eta_V}\\
    \xi_V^2(\phi) &=
    \frac{\al^2 M_{\rm Pl}^6}{\phi^8}
    \left(
        \al^2 M_{\rm Pl}^2
        - 6\al M_{\rm Pl}\phi
        + 6\phi^2
    \right).
\end{align}

Inflation ends when $\epsilon_H\equiv-\dot H/H^2=1$, equivalently $\ddot a=0$. In the slow-roll approximation $\epsilon_H\simeq\epsilon_V$, and we use $\epsilon_V=1$ as an analytic estimate of the end of inflation. This gives the field values at the end of inflation ($\phi_{\rm end}$) and at horizon crossing of the pivot scale ($\phi_\star$) as
\begin{equation}
\phi_{\rm end}
= \frac{\sqrt{\alpha}}{2^{1/4}} M_{\rm Pl},
~~~
\phi_\star
= M_{\rm Pl}\left(3\alpha N_\star + \left(\frac{\alpha^2}{2}\right)^{3/4}\right)^{1/3}.
\label{eq:phi_end_star}
\end{equation}
where $N_\star$ is the number of $e$-folds, $N_\star = \ln(a_{\rm end}/a_\star)$, with $a$ the scale factor.

Evaluated at $\phi_\star$, the inflationary observables at leading order in
slow roll are
\begin{align}
    r &= \frac{8\al^2 M_{\rm Pl}^4}{\phi_\star^4},~~~~
    n_s =  1 - \frac{\al^2 M_{\rm Pl}^4}{\phi_\star^4}
      - \frac{4\al M_{\rm Pl}^3}{\phi_\star^3}, 
      \label{eq:r_ns}\\
    \alpha_s &=  - \frac{4 \alpha^2 M_{\rm Pl}^6 (\alpha M_{\rm Pl} + 3 \phi_\star)}{\phi_\star^7}.
    \label{eq:alpha_s}
\end{align}

Figure~\ref{fig:inflation_parameters} shows the dependence of the inflationary observables, the tensor-to-scalar ratio $r$, the scalar spectral index $n_{\rm s}$, and the running of the scalar spectral index $\alpha_{\rm s}$, on the model parameter $\alpha$. These observables are evaluated at horizon exit for $N_\star=50$–$60$ e-folds. We find that over a wide range of $\alpha$, the model predicts a sufficiently small tensor-to-scalar ratio $r$, while the scalar spectral index remains clustered around $n_{\rm s} \simeq 0.97$, consistent with current observational constraints. The lower figure further shows that the running of the scalar spectral index is nearly negligible across the considered parameter range.

\begin{figure}[ht]
\centering
\includegraphics[scale=0.45]{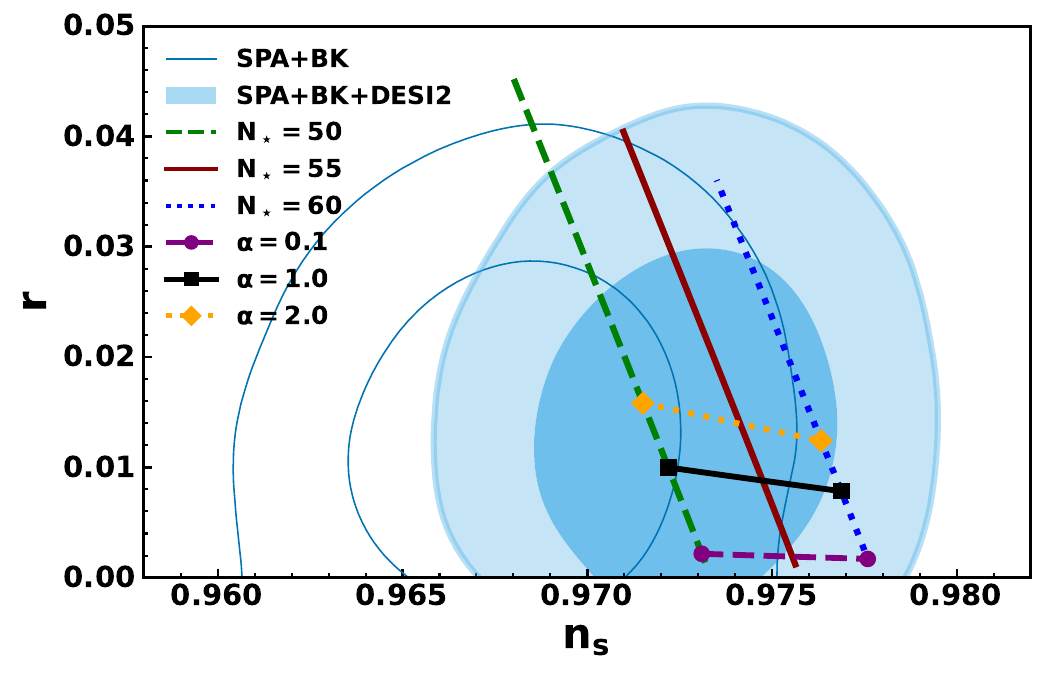}
\caption{Comparison of theoretical predictions for the tensor-to-scalar ratio $r$ and scalar spectral index $n_{\rm s}$ of the IExp potential~\eqref{eq:potential_inflation} with observations for $N_\star=50$--$60$. Shaded contours correspond to the combined SPA+BK+DESI2 data set, while line contours represent SPA+BK constraints, with inner and outer contours denoting the $1\sigma$ and $2\sigma$ regions, respectively. The contours are obtained using publicly available MCMC chains \cite{Balkenhol:2025wms}.}
\label{fig:r-ns}
\end{figure}

In Fig.~\ref{fig:r-ns}, we compare the inflationary predictions of the potential~\eqref{eq:potential_inflation} in the $(n_{\rm s}, r)$ plane with current observational constraints \cite{Planck:2018jri,AtacamaCosmologyTelescope:2025blo,DESI:2025zgx,SPT-3G:2025bzu,BICEP:2021xfz}. The observational contours shown in Fig.~\ref{fig:r-ns} are obtained using publicly available MCMC chains\footnote{\texttt{\url{https://github.com/Lbalkenhol/r_ns_2025}}} \cite{Balkenhol:2025wms}, visualized using the {\tt GetDist} package \citep{Lewis:2019xzd}.

In Fig.~\ref{fig:r-ns} the curves for $N_\star=50$, 55 and 60 show the IExp predictions as $\alpha$ varies over $0.05\le\alpha\le10$, with markers at $\alpha=0.1$, 1 and 2. For $\alpha$ up to a few the predictions lie within the $1\sigma$ region. Increasing $\alpha$ raises $r$ and moves the curves out of this region, and for $\alpha\gtrsim7$--$10$, depending on $N_\star$, $r$ exceeds the bound $r<0.035$. Throughout, $n_{\rm s}$ stays between about $0.968$ and $0.978$.

The overlap between the predicted curves and the observationally preferred region shows that the inverse exponential potential, with the single parameter $\alpha$, is consistent with present observations for $\alpha$ up to a few. The remaining parameter, $V_0$, fixes only the overall energy scale of inflation and does not affect the predictions for $r$ or $n_{\rm s}$.

\begin{figure}[ht]
\centering
\includegraphics[scale=0.45]{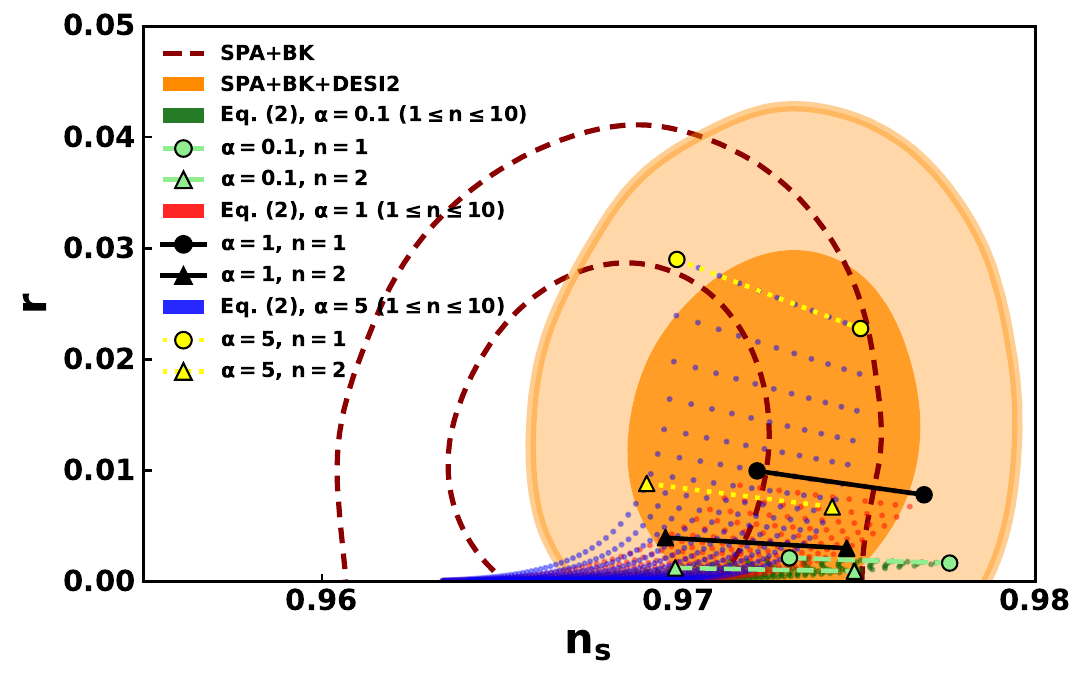}
\caption{Predictions of the generalized potential~\eqref{eq:potential_inflation_general} for $\alpha=0.1,\,1,\,5$, each shown as a band over $-10\le n\le -1$ (at $50\le N_\star\le 60$) together with the endpoint trajectories $n=-1$ (circles, exactly the IExp potential~\eqref{eq:potential_inflation}) and $n=-10$ (triangles). Contours and MCMC chains as in Fig.~\ref{fig:r-ns}.}
\label{fig:r-ns-genn}
\end{figure}
 
In addition to Fig.~\ref{fig:r-ns} for the IExp potential~\eqref{eq:potential_inflation}, we also show the predictions of the generalised IExp potential~\eqref{eq:potential_inflation_general} in Fig.~\ref{fig:r-ns-genn} \footnote{The expressions of the inflationary observables for the potential~\eqref{eq:potential_inflation_general} are given in Ref.~\cite{Geng:2015fla}.}. At each fixed $\alpha$, the prediction moves only mildly as $n$ is decreased from -1 to -10, $n_{\rm s}$ drifts down by a few $\times 10^{-3}$ at fixed $N_\star$, while $r$ falls rapidly, by roughly two orders of magnitude. Over the full range $-10\le n\le -1$ and $\alpha\in\{0.1,1,5\}$, almost every point remains within the $1\sigma$ SPA+BK+DESI2 contour, with $n_{\rm s}$ confined to $0.963\lesssim n_{\rm s}\lesssim 0.978$ and $r\lesssim 0.03$. The observational viability of Eq.~\eqref{eq:potential_inflation_general} for $n<0$ is therefore not a special property of $n=-1$: it persists across at least an order of magnitude in $|n|$, with $n=-1$ singled out only by its simplicity, a single free parameter besides $V_0$, rather than by any special tuning of the $(n_{\rm s},r)$ prediction.

\section{Comparison with other models}
\label{sec:comparison}

In this section we compare the IExp potential~\eqref{eq:potential_inflation} against several other models, the $R^2$ (Starobinsky)~\cite{Starobinsky:1980te,Starobinsky:1983zz}
and monomial potentials~\cite{Linde:1981mu,Linde:1983gd}, $\alpha$-attractor E- and T-models~\cite{Kallosh:2013hoa,Kallosh:2013yoa}, non-minimally coupled Higgs-like
inflation~\cite{Spokoiny:1984bd,Salopek:1988qh,Bezrukov:2007ep}, and a radiatively-corrected plateau potential~\cite{Wolf:2025ecy}.
In every figure below, the shaded band and $\alpha=0.1,1,2$ trajectories reproduce the IExp
comparison band of Fig.~\ref{fig:comparison} ($0.05\le\alpha\le30$, $50\le N_\star\le60$), so that each
model family can be read against the same reference.

\subsection{Starobinsky and monomial potentials}
\label{sec:star-linde}

The $R^2$ (Starobinsky) prediction~\cite{Starobinsky:1980te,Mukhanov:1981xt,Starobinsky:1983zz} is
the large-$N$ attractor limit
\begin{equation}
n_s = 1-\frac{2}{N_\star}, \qquad r=\frac{12}{N_\star^2}.
\end{equation}
Monomial potentials~\cite{Linde:1981mu,Linde:1983gd}, $V(\phi)=V_0\phi^n$, give
\begin{equation}
\epsilon_V=\frac{n^2\Mpl^2}{2\phi^2}, \qquad \eta_V=\frac{n(n-1)\Mpl^2}{\phi^2},
\end{equation}
with $\phi_{\rm end}=n/\sqrt2\Mpl$ and $\phi_\star=\sqrt{2nN_\star\Mpl^2+\phi_{\rm end}^2}$, so that
\begin{equation}
n_s = 1-\frac{n+2}{2N_\star}, \qquad r=\frac{4n}{N_\star}.
\end{equation}
We show the monomial family as a continuous shaded band over $0.01\le n\le5$ at $N_\star=50$--60,
together with the three representative trajectories $\phi$, $\phi^{2/3}$, $\phi^{1/3}$.

\begin{figure}[t]
\centering
\includegraphics[scale=0.45]{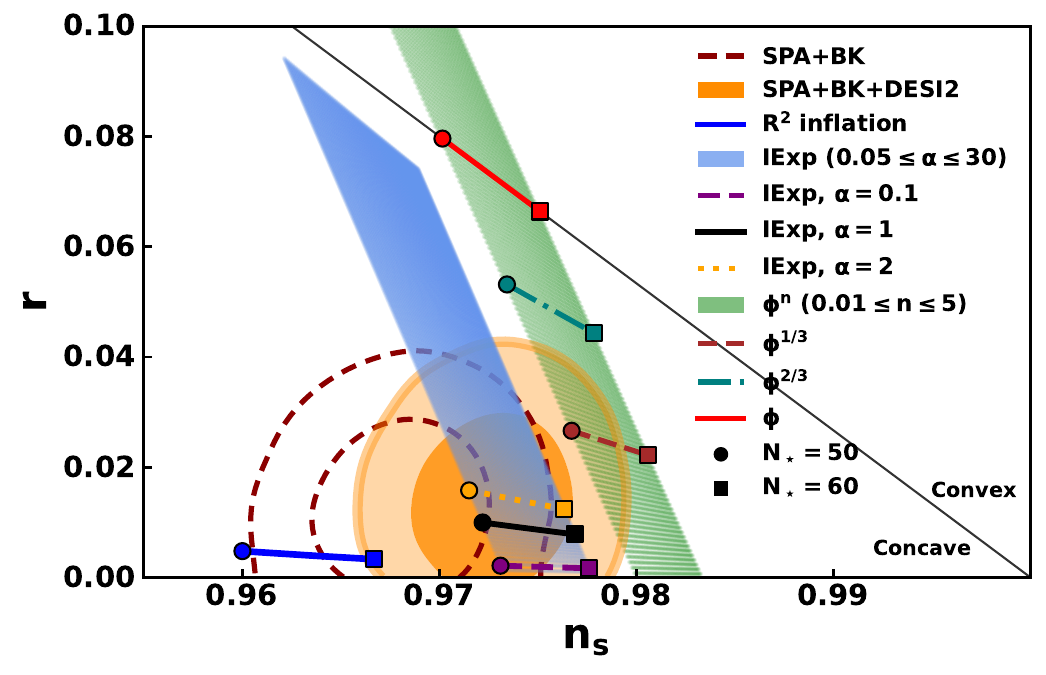}
\caption{Predictions of the IExp potential~\eqref{eq:potential_inflation} together with the $R^2$
model and the monomial potentials $V\propto\phi^n$, shown as a band over $0.01\le n\le5$ together with
the representative trajectories $\phi$, $\phi^{2/3}$, $\phi^{1/3}$, for $N_\star=50$--$60$.
Shaded/line contours and MCMC chains as in Fig.~\ref{fig:r-ns}.}
\label{fig:comparison}
\end{figure}

As shown in Fig.~\ref{fig:comparison}, the $R^2$ model lies near the boundary of the $2\sigma$
SPA+BK+DESI2 region, and the monomial band lies partially or entirely outside it across essentially
the whole $0.01\le n\le5$ range sampled, in contrast to the IExp band for $\alpha$ up to a few, which lies within
$1\sigma$.

\subsection{$\alpha$-attractor E- and T-models}

We compare against the canonical $\alpha$-attractor family~\cite{Kallosh:2013hoa,Kallosh:2013yoa},
recently re-examined against ACT DR6~\cite{Ellis:2025zrf,Haque:2025uri}. The E- and T-model potentials are
\begin{align}
V_E(\phi) &= V_0\left[1-\exp\!\left(-\sqrt{\tfrac{2}{3\alpha}}\,\frac{\phi}{\Mpl}\right)\right]^2, \\
V_T(\phi) &= V_0\tanh^2\!\left(\frac{\phi}{\sqrt{6\alpha}\Mpl}\right),
\end{align}
with $\phi$ already the canonically normalized field, so the standard slow-roll expressions
$\epsilon_V=\tfrac12(V'/V)^2$, $\eta_V=V''/V$, $n_s=1-6\epsilon_V+2\eta_V$, $r=16\epsilon_V$ apply directly. We show each model as a band over $0.1\le\alpha\le10$ at $N_\star=50$--60, together with the representative trajectories $\alpha=1,4,10$.

\begin{figure}[t]
\centering
\includegraphics[scale=0.45]{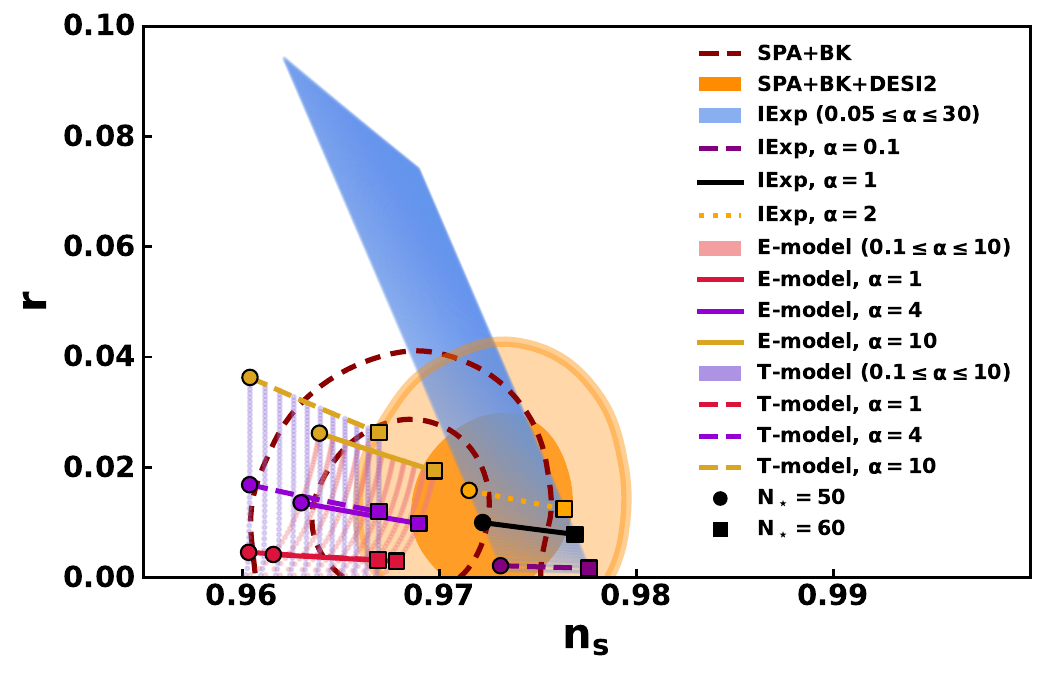}
\caption{IExp potential~\eqref{eq:potential_inflation} compared to the E- and T-model
$\alpha$-attractors, each shown as a band over $0.1\le\alpha\le10$ (at $N_\star=50$--60) together with
representative trajectories $\alpha=1,4,10$.}
\label{fig:attractors}
\end{figure}

Both models reduce to the Starobinsky prediction as $\alpha\to1$ and to $r\propto\alpha/N_\star^2$ for
general $\alpha$, the defining attractor behaviour of this class. Fig.~\ref{fig:attractors} shows both
bands together with $\alpha=1,4,10$, larger $\alpha$ pushes $r$ up while leaving $n_s$ close to the
Starobinsky value.

\subsection{Non-minimally coupled Higgs-like inflation}
\label{sec:nonmin-higgs}
Following~\cite{Gao:2025onc,Spokoiny:1984bd,Salopek:1988qh,Bezrukov:2007ep}, we consider
$f(\phi)=(\phi/\Mpl)^k$ and couple it to gravity in the Jordan frame (subscript $J$) through
\begin{equation}
\Omega^2(\phi) = 1+\xi f(\phi), \qquad V_J(\phi)=\lambda^2\Mpl^4 f^2(\phi),
\end{equation}
so that for $f(\phi)=(\phi/\Mpl)^k$,
\begin{equation}
\Omega^2(\phi) = 1+\xi\frac{\phi^k}{\Mpl^k}, \qquad V_J(\phi)=\lambda^2\Mpl^4\frac{\phi^{2k}}{\Mpl^{2k}}\, .
\label{eq:nonmin-Om_VJ}
\end{equation}
This gives the Einstein-frame potential $U(\phi)=V_J(\phi)/\Omega^4(\phi)$ and a non-canonical field-space
metric
\begin{equation}
K(\phi)\equiv\left(\frac{d\chi}{d\phi}\right)^2 = \frac{1}{\Omega^2}+\frac{3}{2}\Mpl^2\left(\frac{d\Omega^2/d\phi}{\Omega^2}\right)^2 \, .
\end{equation}
The slow-roll parameters generalize to
\begin{equation}
\epsilon_V=\frac{U'^2}{2KU^2}, \qquad
\eta_V=\frac{U''}{KU}-\frac{U'K'}{2K^2U},
\end{equation}
with $N_\star=\int_{\phi_{\rm end}}^{\phi_\star}(KU/U')\,d\phi$. For $k=2$, Eq.~\eqref{eq:nonmin-Om_VJ} reduces to the non-minimally coupled quartic potential
$\Omega^2(\phi)=1+\xi\phi^2/\Mpl^2$, $V_J(\phi)=\lambda^2\phi^4$, the standard Higgs-inflation
construction of Refs.~\cite{Spokoiny:1984bd,Salopek:1988qh,Bezrukov:2007ep}. Ref.~\cite{Gao:2025onc} constrains this construction against current observations to
$0\lesssim k\lesssim1.5$ and $0.1\lesssim\xi\lesssim35$. The standard Higgs case, $k=2$, lies outside
this range and is therefore observationally disfavoured within this construction. Rather than sampling
isolated $(k,\xi)$ points, we fix two representative values inside the viable window, $k=\tfrac13$ and
$k=\tfrac23$, and for each scan $\xi$ continuously over the full reported range $0.1\le\xi\le35$ at
$N_\star=50$--60, together with the trajectories $\xi=1$ and $\xi=20$ marking the low and high ends of
that range.

\begin{figure}[t]
\centering
\includegraphics[scale=0.45]{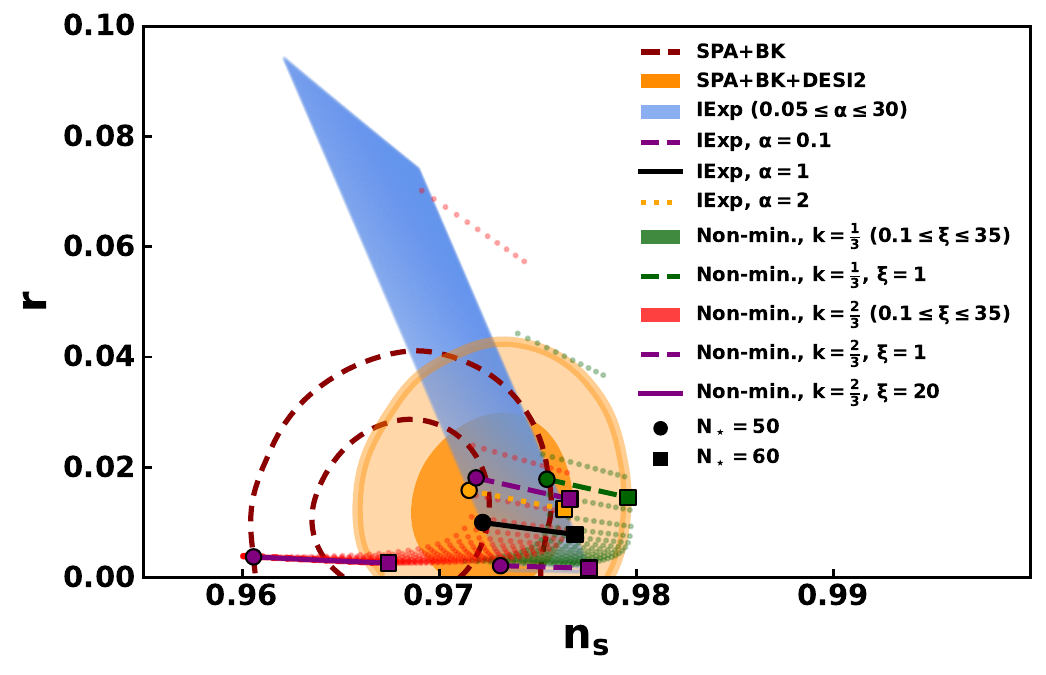}
\caption{IExp potential~\eqref{eq:potential_inflation} compared to the non-minimally coupled
generalized-Higgs family for $k=1/3$ and $k=2/3$, each shown as a band over $0.1\le\xi\le35$ (at
$N_\star=50$--60) together with representative trajectories $\xi=1,20$.}
\label{fig:nonminimal}
\end{figure}

Fig.~\ref{fig:nonminimal} shows both bands, increasing $\xi$ at fixed $k$ suppresses $r$ and pulls
$n_s$ down toward the Starobinsky value $n_s=1-2/N_\star$, consistent with the general tendency of
non-minimally coupled monomial constructions to approach a common plateau-type attractor as the
coupling strengthens. Both the $k=1/3$ and $k=2/3$ bands remain comfortably within the region already
favoured by SPA+BK+DESI2 across the entire range of $\xi$.

\subsection{Radiatively-corrected (quantum-corrected) plateau}
Following~\cite{Wolf:2025ecy}, we consider a one-loop-motivated logarithmically running quartic coupling on top of the $k=2$ non-minimally coupled construction of Sec.~\ref{sec:nonmin-higgs},
\begin{align}
\Omega^2(\phi) &= 1+\xi\left(\frac{\phi}{\Mpl}\right)^2, \\
\lambda(\phi) &= 1+\beta\ln\left(\frac{\phi}{\Mpl}\right),
\\ V_J(\phi) &= \lambda(\phi)\,\left(\frac{\phi}{\Mpl}\right)^4,
\end{align}
with $U(\phi)=V_J(\phi)/\Omega^4(\phi)$ and $K(\phi)$ as in Sec.~\ref{sec:nonmin-higgs} at $k=2$. This is the $\xi$-attractor construction of Ref.~\cite{Wolf:2025ecy}, with $\beta$ playing the role of radiative correction strength. We scan $\xi$ over $0.1\le\xi\le300$ at fixed $\beta=0.01,0.02$ (at $N_\star=50$--60), together with the representative trajectories $\xi=1,10,100$ for each.

\begin{figure}[t]
\centering
\includegraphics[scale=0.45]{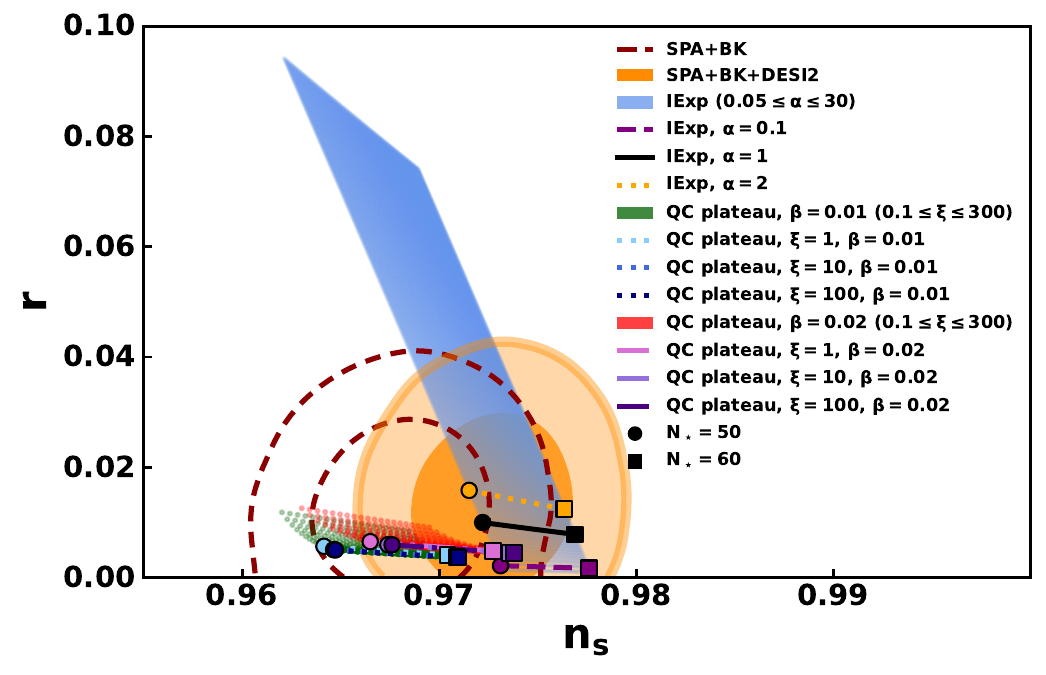}
\caption{IExp potential~\eqref{eq:potential_inflation} compared to the radiatively-corrected plateau
potential, shown as bands over $0.1\le\xi\le300$ at fixed $\beta=0.01$ and $\beta=0.02$, together with
representative trajectories $\xi=1,10,100$ at each $\beta$.}
\label{fig:qcplateau}
\end{figure}

Fig.~\ref{fig:qcplateau} shows both $\beta$ bands across the full $0.1\le\xi\le300$ range. At fixed $\beta$, the predicted $(n_s,r)$ shifts only mildly, remaining within the SPA+BK+DESI2 $1\sigma$ region for larger $N_\star$ values for both values of $\beta$. Doubling $\beta$ from $0.01$ to $0.02$ shifts each trajectory only mildly within its own band, indicating that the effect on $(n_s,r)$ is controlled primarily by $\xi$ at fixed percent-level
$\beta$.

\section{Post-inflationary dynamics}
\label{sec:reheating}

\begin{figure}[t]
\centering
\includegraphics[scale=0.45]{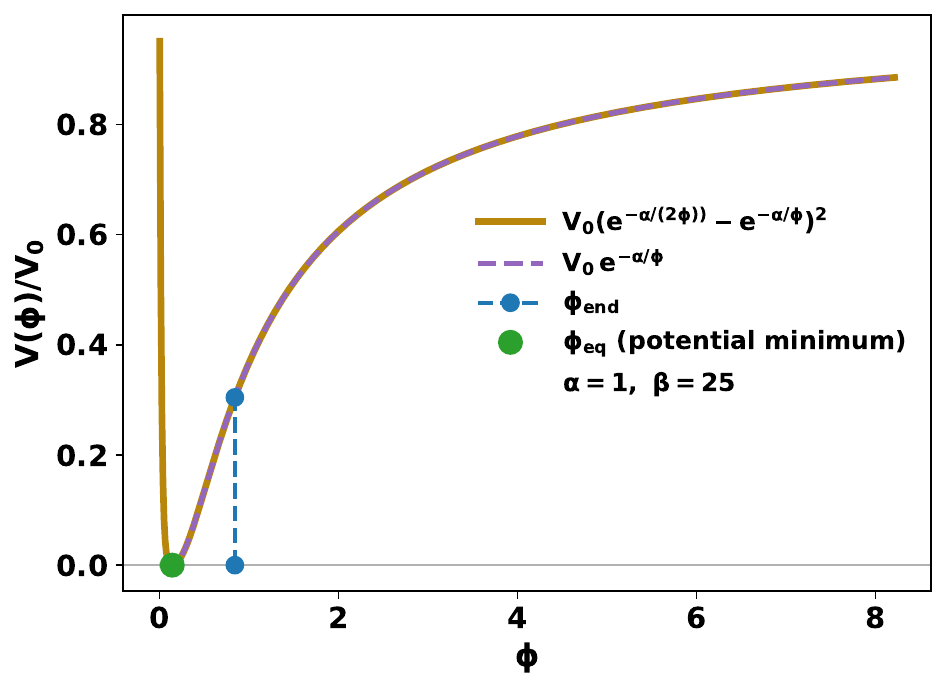}
\caption{Inflationary potentials~\eqref{eq:potential_inflation} and \eqref{eq:Vsquared} for $\alpha=1$ and $\beta=25$.}
\label{fig:potential_full}
\end{figure}

To ensure a graceful exit from inflation and a consistent reheating
phase, we modify the inflationary potential~\eqref{eq:potential_inflation} to
\begin{equation}
V(\phi) = V_0\Big(e^{-\alpha M_{\rm Pl}/(2\phi)} - e^{-\beta\phi/M_{\rm Pl}}\Big)^2,
\qquad \alpha,\beta>0\, .
\label{eq:Vsquared}
\end{equation}
During inflation, $\beta\phi/M_{\rm Pl}\gg1$ and Eq.~(\ref{eq:Vsquared}) reduces to
\begin{equation}
V(\phi) \;\longrightarrow\; V_0\,e^{-\alpha\Mpl/\phi},
\end{equation}
i.e.\ exactly the IExp potential of Eq.~\eqref{eq:potential_inflation} as shown in Fig.~\ref{fig:potential_full}.

Because Eq.~(\ref{eq:Vsquared}) is a perfect
square, $V(\phi)\geq0$ for all $\phi>0$ automatically, with equality if and
only if the two exponentials coincide at $\phi_{\rm eq}$, precisely the condition
$e^{-\alpha\Mpl/(2\phi_{\rm eq})}=e^{-\beta\phi_{\rm eq}/\Mpl}$, which gives
\begin{eqnarray}
    \phi_{\rm eq}=\sqrt{\frac{\alpha}{2\beta}}\Mpl \, .
\end{eqnarray}
For the modified potential~\eqref{eq:Vsquared}, $V'(\phi_{\rm eq})=0$ and the curvature at the minimum,
\begin{equation}
V''(\phi_{\rm eq}) = 8V_0\Mpl^{-2}\,\beta^2\,e^{-\sqrt{2\alpha\beta}},
\label{eq:Vpp_sq}
\end{equation}
which is positive, 
confirming $\phi_{\rm eq}$ is a genuine minimum.

The condition for a graceful exit, $\phi_{\rm eq}<\phi_{\rm end}$, reads
$\sqrt{\alpha/(2\beta)} < \sqrt{\alpha}/2^{1/4}$, i.e.
\begin{equation}
\beta > \frac{1}{\sqrt{2}} \approx 0.71.
\end{equation}

This algebraic bound only guarantees that the two exponentials in
Eq.~(\ref{eq:Vsquared}) cross after the field reaches $\phi_{\rm end}$.
The curvature of the potential at the minimum, given by Eq.~(\ref{eq:Vpp_sq}), determines the effective inflaton mass, $m_\phi^2\equiv V''(\phi_{\rm eq})$. At the horizon crossing of the pivot scale $ k_\star = 0.05~{\rm Mpc}^{-1}$, the Hubble parameter is $H_\star = \pi \sqrt{r A_{\rm s}(k_\star)/2}M_{\rm Pl}\approx \sqrt{V(\phi_\star)/(3\Mpl^2)}$. Using the observed amplitude of the scalar power spectrum at horizon crossing, $A_{\rm s}(k_\star) = 2.10 \times 10^{-9}$ \cite{Planck:2018jri}, we can write the inflationary energy scale as
\begin{align}
V(\phi_\star)^{1/4} &= 1.33\times 10^{-2} r^{1/4} \Mpl \, , \nn \\  &=3.233 \times 10^{16} r^{1/4}~{\rm GeV},
\label{eq:Vscale_general}
\end{align}
which can be used to determine $V_0$ for a particular $r$. For $\alpha=1$ and $N_\star=55$ the model gives $r\simeq0.009$, and we adopt the representative value $r=0.01$, for which $V_0^{1/4}=1.07\times 10^{16}~{\rm GeV}$. This gives
$m_\phi\approx 1.3\times10^{14}$, $9.7\times10^{13}$, and
$8.3\times10^{13}\,{\rm GeV}$ for $\beta=15,25,30$ respectively. $m_\phi$
falls with increasing $\beta$ because the exponential suppression
$e^{-\sqrt{2\alpha\beta}}$ in Eq.~(\ref{eq:Vpp_sq}) outweighs the explicit
$\beta^2$ prefactor once $\alpha\beta\gtrsim 8$. 

As the field rolls toward the minimum, however, the term $e^{-\beta\phi/\Mpl}$ becomes increasingly important and governs the post-inflationary dynamics. To check this quantitatively for a given $\alpha$, we solve the full Klein--Gordon equation,
$\ddot\phi+3H\dot\phi+V'(\phi)=0$, numerically from the end of inflation
onward. Since the detailed physics of the reheating era is not specified by
the model, we take the oscillating scalar field to be the sole component of
the Universe during this phase, as is standard in this type of estimate.

\begin{figure}[t]
\centering
\includegraphics[scale=0.45]{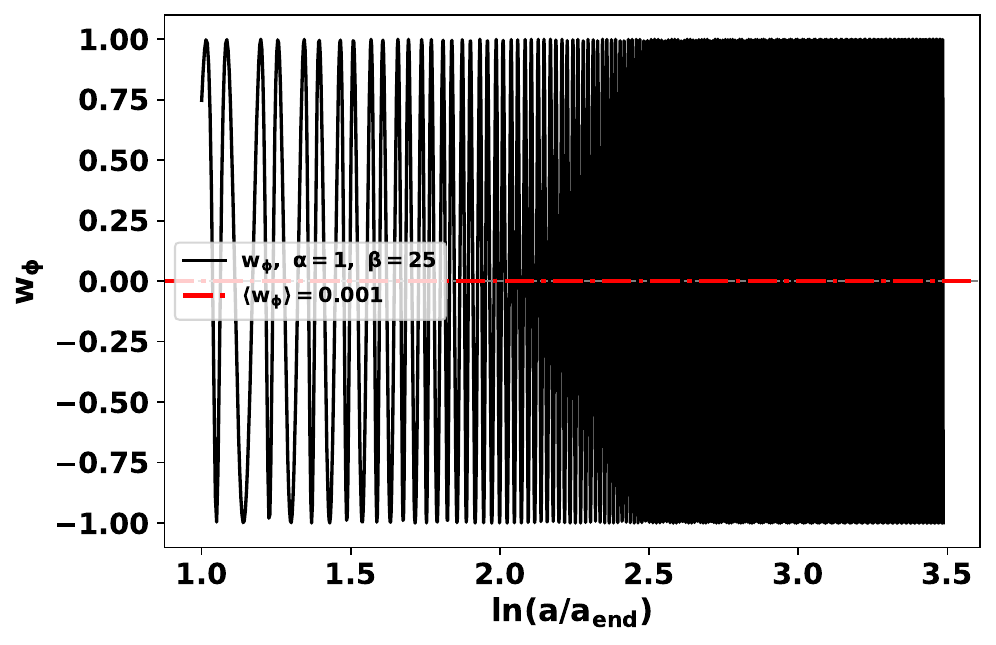}\\
\includegraphics[scale=0.45]{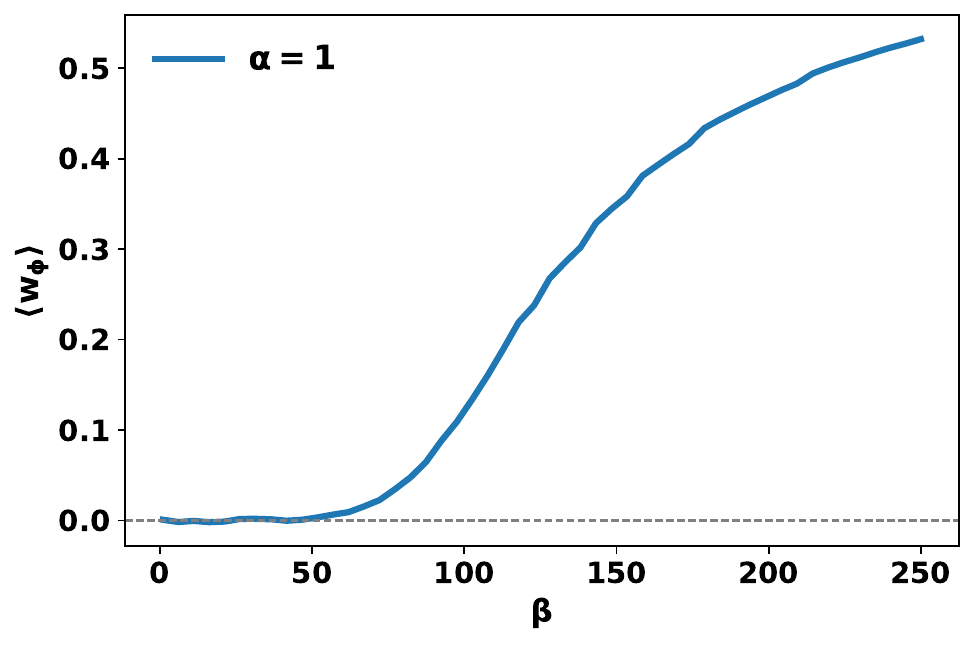}
\caption{Post-inflationary evolution of the scalar field for the potential~\eqref{eq:Vsquared}. {\it Top:} evolution of $w_\phi$ for $\alpha=1$ and $\beta=25$. Red dashed line indicates the time-averaged value. {\it Bottom:} time-averaged EoS $\langle w_\phi \rangle$ as a function of $\beta$ for $\alpha=1$.}
\label{fig:post-inf}
\end{figure}

Once inflation ends, the potential minimum allows the scalar field to
oscillate, driving reheating. The top panel of Fig.~\ref{fig:post-inf} shows the resulting evolution of the field's equation of state (EoS) $w_\phi$ as a function of e-folds from the end of inflation, for $\alpha=1$ and $\beta=25$. The red dashed line marks its time-averaged value. The bottom
panel shows this time-averaged EoS $\langle w_\phi\rangle$ as a function of
$\beta$ for $\alpha=1$, over $1\leq\beta\leq250$. $\langle w_\phi\rangle$ is close to zero for $\beta\lesssim30$, consistent with the near-harmonic oscillation expected about a quadratic minimum, and rises monotonically to $\langle w_\phi\rangle\approx0.2$ at $\beta\approx115$ and $\approx0.5$ at $\beta=250$ as the steepness of the exponential potential increasingly stiffens the oscillations away from the
minimum. This behaviour arises because the minimum of the potential in Eq.~\eqref{eq:Vsquared} becomes progressively broader as $\beta$ increases, analogous to the $\phi^n$ potential, whose minimum also widens with increasing values of $n$. Across this entire range $\beta$ lies far above the $\beta>1/\sqrt2$
threshold derived above, so the field oscillates cleanly about $\phi_{\rm
eq}$ with no residual accelerated phase, in agreement with the analytic
graceful-exit condition.

\section{Reheating}
\label{sec:Trh}

\subsection{Reheating temperature}
\label{sec:Trh-bookkeeping}

\begin{figure*}[t]
\centering
\includegraphics[scale=0.35]{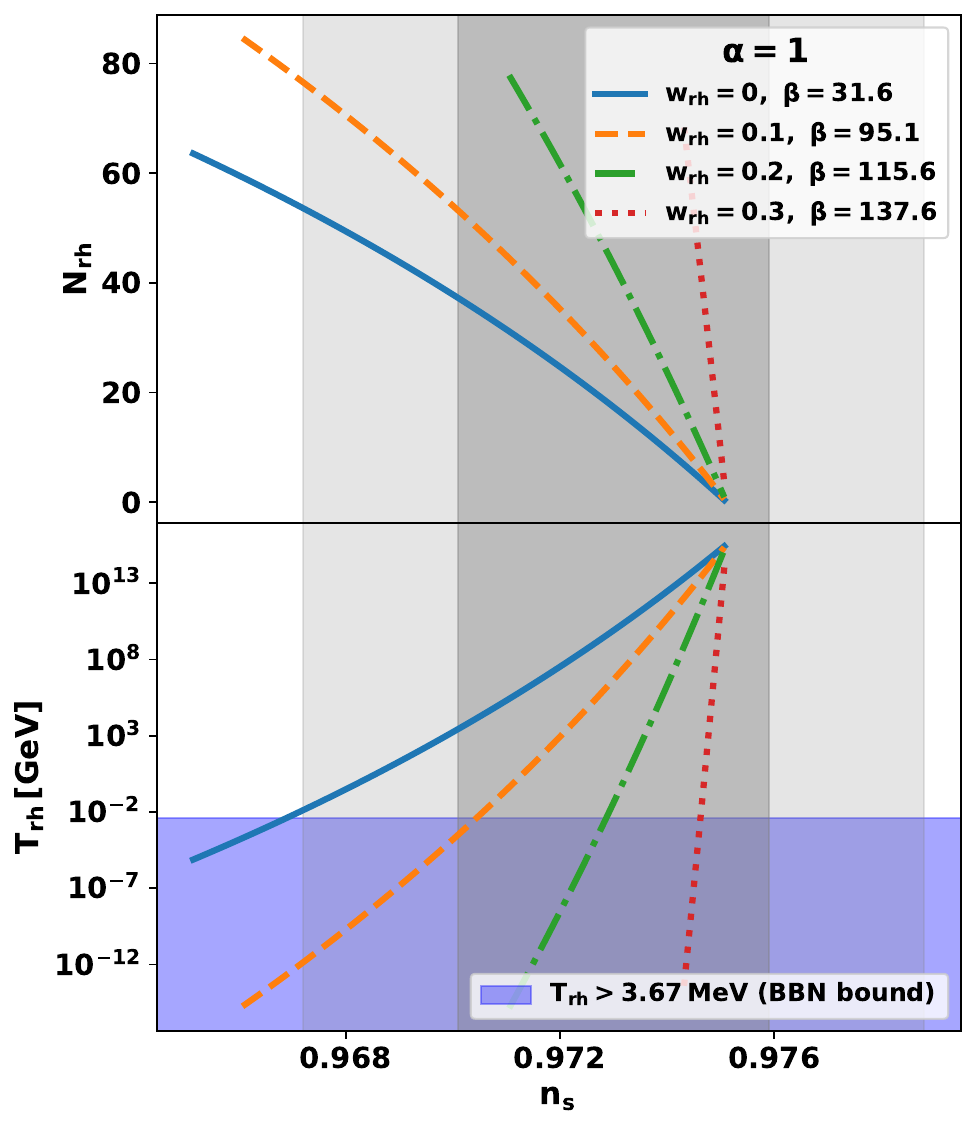}
\includegraphics[scale=0.35]{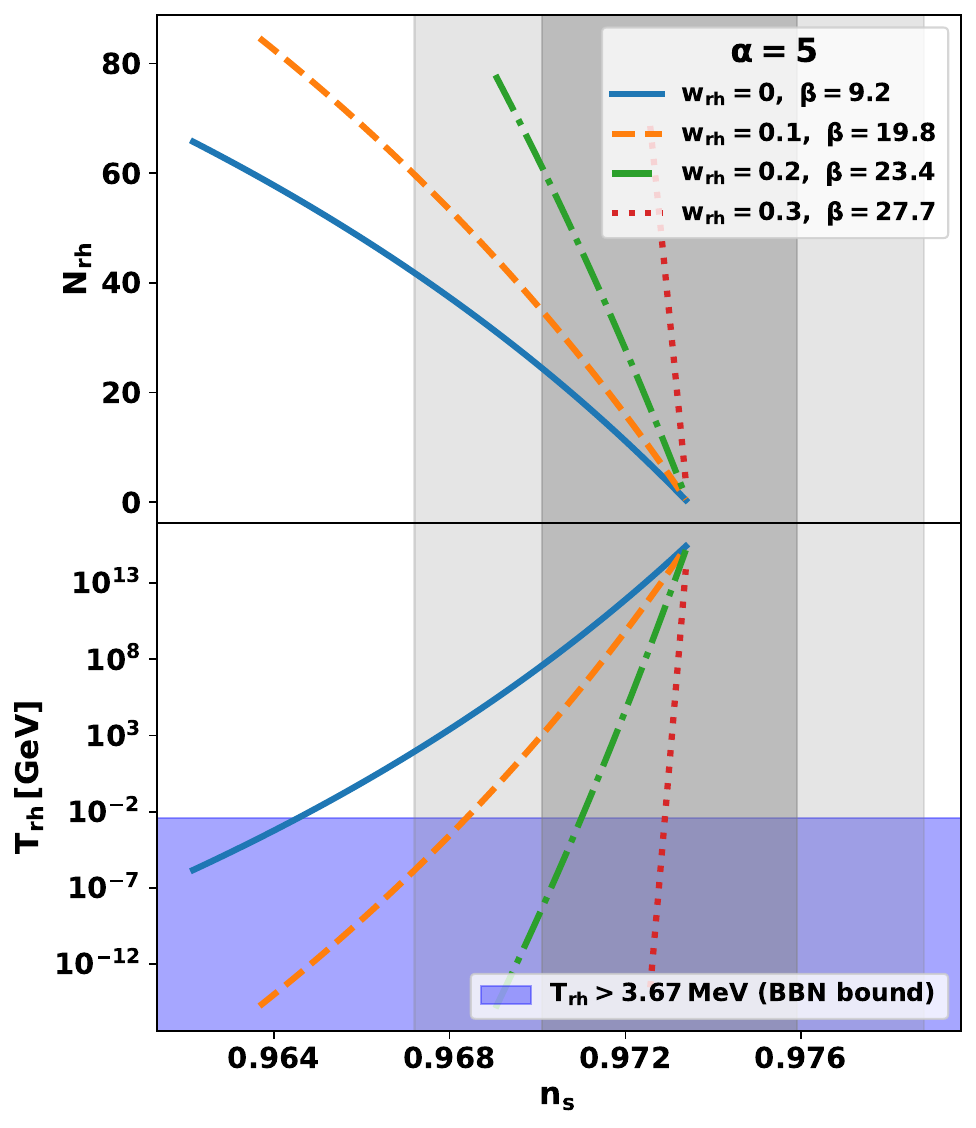}\\
\includegraphics[scale=0.35]{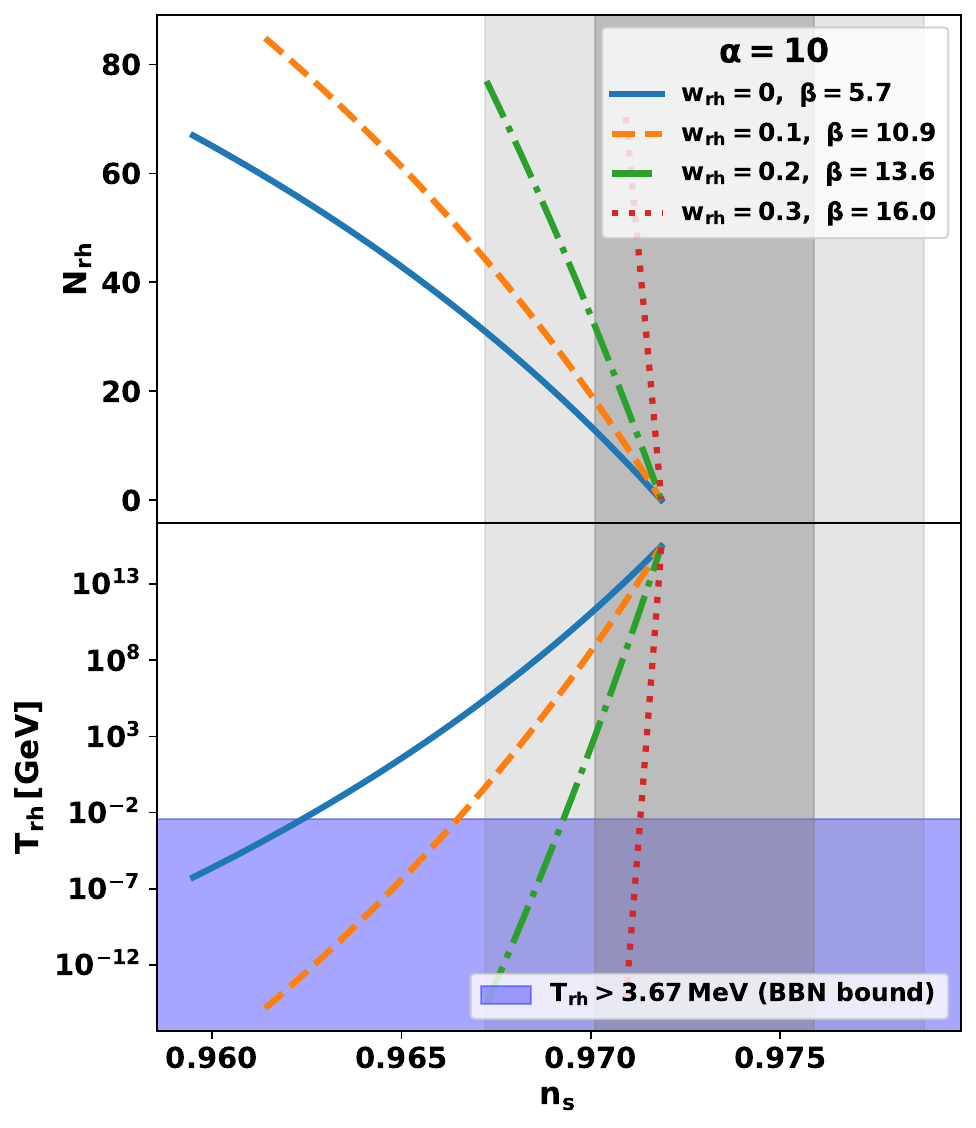}
\includegraphics[scale=0.35]{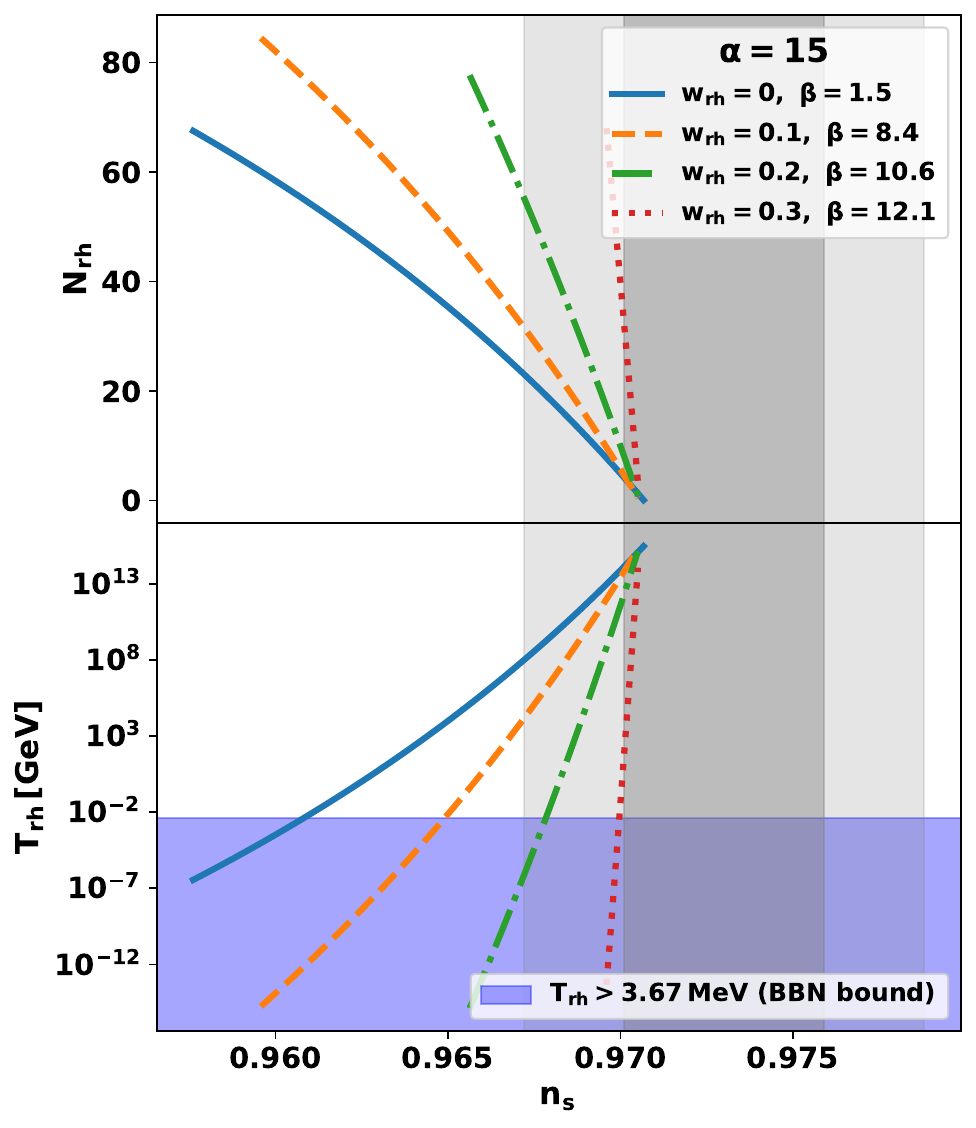}
\caption{Reheating e-folding number and reheating temperature predicted by the potential~\eqref{eq:Vsquared}, for $\alpha=1,5,10,15$. In each panel, $\beta$ is not fixed but chosen separately for every curve so that the time-averaged equation of state obtained from the numerical Klein--Gordon evolution of Sec.~\ref{sec:reheating} matches the assumed reheating equation of state, $w_{\rm rh}=0,0.1,0.2,0.3$ (the corresponding $\beta$ is quoted in each legend). The shaded bands mark the $1\sigma$ and $2\sigma$ SPA+BK+DESI2 range of $n_s$, and the horizontal band marks the region excluded by the BBN bound $T_{\rm rh}>3.67\,{\rm MeV}$.}
\label{fig:reheating}
\end{figure*}

After inflation, the inflaton field transfers its energy density to relativistic particles, initiating reheating \cite{Kofman:1994rk,Allahverdi:2010xz,Lozanov:2019jxc}. The reheating temperature, $T_{\rm rh}$, is defined as the temperature of the thermal bath at the end of reheating when the Universe becomes radiation-dominated with corresponding energy density $\rho_{\rm rh}$. Assuming instantaneous thermalisation, it can be written as \cite{Cook:2015vqa},
\begin{eqnarray}
T_{\rm rh} &=& \left(\frac{30 \rho_{\rm rh}}{\pi^2 g_{\rm rh}}\right)^{1/4} \, , 
\label{eq:Trh}\\
\rho_{\rm rh} &=& \rho_{\rm end} \left(\frac{a_{\rm end}}{a_{\rm rh}}\right)^{3(1+w_{\rm rh})}\, ,
\label{eq:Treh_def}
\end{eqnarray}
where $g_{\rm rh}$ is the effective number of relativistic degrees of freedom at the end of reheating, at scale factor $a_{\rm rh}$. Here, $\rho_{\rm rh}$ is related to the energy density at the end of inflation, $\rho_{\rm end}=\tfrac32 V_{\rm end}$ (with $V_{\rm end}$ evaluated on the potential of Eq.~\eqref{eq:Vsquared}), assuming reheating proceeds with a constant effective EoS parameter $w_{\rm rh}$.

The reheating e-fold number, $N_{\rm rh}$, and the reheating temperature, $T_{\rm rh}$, can be expressed in terms of inflationary observables as \cite{Cook:2015vqa} (see Appendix~\ref{app:reheat} for details)
\begin{eqnarray}
N_{\rm rh} &=&
\frac{4}{1-3w_{\rm rh}} \left( 61.6 - \ln\!\left(\frac{V_{\rm end}^{1/4}}{H_\star}\right) - N_\star \right),
\label{eq:Nrh} \\
T_{\rm rh} &=& \left(\frac{45V_{\rm end}}{\pi^2 g_{\rm rh}}\right)^{1/4}
\exp\Big[-\frac{3}{4}(1+w_{\rm rh})\,N_{\rm rh}\Big] \,. \nn \\
&=& \Bigg\{ \left(\frac{43}{11 g_{\rm rh}}\right)^{1/3}\frac{a_0 T_0}{k_\star} H_\star e^{-N} \nn \\ && \times\left(\frac{45 V_{\rm end}}{\pi^2 g_{\rm rh}}\right)^{- \frac{1}{3(1+w_{\rm rh})}} \Bigg\}^{\frac{3(1+w_{\rm rh})}{3w_{\rm rh}-1}},
\label{eq:Trh_standard}
\end{eqnarray}
valid for $w_{\rm rh}\neq 1/3$ and from Eq.~\eqref{eq:Vscale_general} we have $V(\phi_\star)^{1/4} \simeq 1.02 \times 10^{16}~{\rm GeV}$ for $r = 0.01$. The subtracted constant in Eq.~\eqref{eq:Vsquared} is negligible at $\phi_\star$, so this normalization is unaffected by the vacuum-energy subtraction; it does, however, affect $V_{\rm end}$, and hence $N_{\rm rh}$ and $T_{\rm rh}$ below, at the level of a few percent for the $\beta$ values used here.

Figure~\ref{fig:reheating} shows the reheating e-folding number and reheating temperature for four values of the effective equation of state, $w_{\rm rh}=0,0.1,0.2,0.3$, at four representative values of $\alpha$. Rather than fixing $\beta$ and treating $w_{\rm rh}$ as an independent free parameter, here $\beta$ is calibrated separately for each $(\alpha,w_{\rm rh})$ pair so that the time-averaged equation of state $\langle w_\phi\rangle$ obtained by numerically solving the Klein--Gordon equation, as in Sec.~\ref{sec:reheating}, equals the quoted $w_{\rm rh}$. This ties the assumed reheating equation of state self-consistently to the potential parameter $\beta$ rather than leaving it as a free input, and the required $\beta$ decreases with increasing $\alpha$ at fixed $w_{\rm rh}$ and increases with $w_{\rm rh}$ at fixed $\alpha$, consistent with the $\beta$-dependence of $\langle w_\phi\rangle$ found in Fig.~\ref{fig:post-inf}. As $\alpha$ increases, the allowed region shifts toward smaller values of the scalar spectral index $n_s$. Nevertheless, in all cases, the predicted reheating temperatures remain consistent with the observationally allowed $1\sigma$ range of $n_s$.

The maximum reheating temperature predicted by the model typically lies in the range $T_{\rm rh} \sim 10^{12}$–$10^{13}\,{\rm GeV}$. On the lower end, the minimum reheating temperature compatible with the $1\sigma$ constraint on $n_s$ depends on $\alpha$, reaching as low as $< 10^{-4}\,{\rm GeV}$ for $\alpha = 1$ and $5$, while increasing to a few GeV for larger $\alpha$ within the same $1\sigma$ bound. In general, lower reheating temperatures are allowed for smaller $\alpha$ and larger $w_{\rm rh}$, whereas higher temperatures are preferred within the observationally allowed parameter space. We caution that the low-temperature end of this range is not automatically physical. A successful Big Bang Nucleosynthesis requires, at the 95\% confidence level, $T_{\rm rh}\gtrsim$ 3.67 MeV \cite{Barbieri:2025moq}, well above $10^{-4}\,{\rm GeV}=0.1\,{\rm MeV}$, so the corresponding region of parameter space should be regarded as excluded rather than merely disfavoured, notwithstanding its formal consistency with the $1\sigma$ range of $n_s$.

The values of $T_{\rm rh}$ and $N_{\rm rh}$ derived above follow from energy-density alone, with $w_{\rm rh}$ treated as a free effective parameter. They do not by themselves establish that the inflaton actually decays efficiently enough to reach the quoted temperature, nor that the coupling required to do so leaves the flatness of the inflationary potential unspoiled by radiative corrections. We address both points explicitly in the rest of this section: Sec.~\ref{sec:yukawa} works out a minimal perturbative decay channel and solves, in closed form, for the coupling strength needed to reproduce a given $T_{\rm rh}$; Sec.~\ref{sec:radstab} then computes the one-loop correction that this same coupling induces on the inflationary potential and derives, again in closed form, the reheating temperature above which that correction becomes comparable to the tree-level potential.

\subsection{A minimal Yukawa completion}
\label{sec:yukawa}

\subsubsection{Effective mass at the minimum}

Near $\phi_{\rm eq}=\sqrt{\alpha/(2\beta)}\,M_{\rm Pl}$, the potential of Eq.~\eqref{eq:Vsquared} is well approximated by the harmonic form
\begin{equation}
V(\phi) \simeq V(\phi_{\rm eq}) + \tfrac12 m_\phi^2 (\phi-\phi_{\rm eq})^2,
\label{eq:harmonic}
\end{equation}
with the curvature $m_\phi^2 \equiv V''(\phi_{\rm eq})$ given exactly by Eq.~\eqref{eq:Vpp_sq}. Unless stated otherwise, below, we use $\beta=25$ for which $m_\phi\approx 9.7\times10^{13}\,{\rm GeV}$ (Sec.~\ref{sec:reheating}).

\subsubsection{Perturbative decay rate}

We couple the field $\phi$ to a Dirac fermion $\psi$ through a single renormalizable Yukawa interaction. The full Lagrangian for the coupled system is
\begin{equation}
\mathcal{L} = -\frac{1}{2}(\partial\phi)^2 - V(\phi) + \bar\psi\left(i\gamma^\mu\partial_\mu - \lambda_{\rm Y}(\phi-\phi_{\rm eq})\right)\psi \, ,
\label{eq:full_lagrangian}
\end{equation}
where $V(\phi)$ is the potential of Eq.~\eqref{eq:Vsquared}, $\lambda_{\rm Y}$ is a dimensionless Yukawa coupling. We do not consider any bare fermion mass. So the fermion is massless in the absence of the coupling to $\phi$, and its entire mass is field dependent. Considering $\chi\equiv\phi-\phi_{\rm eq}$, the interaction term can be written as
\begin{equation}
\mathcal{L}_{\rm int} = -\lambda_{\rm Y}\,\chi\,\bar\psi\psi \, ,
\label{eq:yukawa_int}
\end{equation}
and the resulting field dependent fermion mass is given by
\begin{equation}
m_\psi(\phi) = \lambda_{\rm Y}(\phi-\phi_{\rm eq}) = \lambda_{\rm Y}\chi \, .
\label{eq:mpsi_def}
\end{equation}
Near the post-inflationary minimum, where $\chi$ oscillates during reheating, $|m_\psi|\to0$ as $\chi\to0$. During inflation, at the horizon crossing, the background field is $\phi=\phi_\star\gg\phi_{\rm eq}$, and the same mass function gives $m_\psi(\phi_\star)=\lambda_{\rm Y}(\phi_\star-\phi_{\rm eq})\simeq\lambda_{\rm Y}\phi_\star$. The small fermion mass near the zero crossings during reheating and the much larger fermion mass during inflation are two evaluations of the same mass function~\eqref{eq:mpsi_def} at two different background values of $\phi$.

Eq.~\eqref{eq:yukawa_int} is the simplest coupling that gives $\chi$ a two-body decay channel. We choose it over a $\chi^2\bar\psi\psi$ or $\chi^2\varphi^2$ coupling for different reasons. A $\chi^2\varphi^2$ coupling to bosons can lead to exponential particle production through parametric resonance \cite{Kofman:1997yn}, so the energy transfer is nonperturbative and is not captured by a simple decay width. A $\chi^2\bar\psi\psi$ coupling to fermions cannot grow exponentially, since Pauli blocking bounds the occupation number \cite{Greene:1998nh,Kofman:1997yn}, but its production is still nonperturbative and is not described by a tree-level rate either.

Near the minimum, $\chi$ obeys the Klein--Gordon equation $\ddot\chi+3H\dot\chi+m_\phi^2\chi=0$. Once $H\ll m_\phi$, this reduces to $\ddot\chi+m_\phi^2\chi=0$, so $\chi$ oscillates as $\chi(t)\approx\Phi\cos(m_\phi t)$, where $\Phi$ is the oscillation amplitude, with frequency $m_\phi$.

The fermion obeys the Dirac equation with this oscillating background mass, $m_\psi(t)=\lambda_{\rm Y}\chi(t)$,
\begin{equation}
\left[i\gamma^\mu\partial_\mu - m_\psi(t)\right]\psi = 0 \, .
\label{eq:dirac_full}
\end{equation}
Following \cite{Greene:1998nh}, we write $\psi=\left[i\gamma^\mu\partial_\mu+m_\psi(t)\right]X$ for an auxiliary field $X(x^\mu)$ and substitute into Eq.~\eqref{eq:dirac_full}. Because $m_\psi(t)$ depends only on time, this gives a decoupled second-order equation for $X$,
\begin{equation}
\Box X - i\gamma^0\dot m_\psi(t) X + m_\psi^2(t) X = 0 \, ,
\label{eq:X_eq}
\end{equation}
where $\Box\equiv\partial^\mu\partial_\mu$. Expanding $X$ in modes, $X\propto f_k(t)\,e^{i\mathbf{k}\cdot\mathbf{x}}R_r$, with $R_r$ an eigenvector of $\gamma^0$ of eigenvalue $+1$, Eq.~\eqref{eq:X_eq} becomes
\begin{equation}
\ddot f_k + \left[k^2 + m_\psi^2(t) - i\dot m_\psi(t)\right] f_k = 0 \, ,
\label{eq:fermion_mode_exact}
\end{equation}
where $k$ is the physical momentum of the fermion mode. The imaginary term drives particle creation at each zero-crossing of $\chi$. A full numerical solution has to keep this term. For locating which momenta are excited, only the real part of the frequency matters, so we define
\begin{align}
&\ddot f_k + \Omega_k^2(t)\, f_k = 0 \, , \\ 
&\Omega_k^2(t) = k^2 + m_\psi(t)^2 \, ,\qquad  m_\psi(t)=\lambda_{\rm Y} \chi(t) \, ,
\label{eq:fermion_mode_eq}
\end{align}
which sets the frequency scale for the fermion modes. Because the effective mass $\lambda_{\rm Y}\chi(t)$ oscillates particle production is governed not by a decay width but by how adiabatically $\Omega_k(t)$ changes. Modes for which
\begin{equation}
\left|\frac{\dot\Omega_k}{\Omega_k^2}\right| \lesssim 1
\label{eq:nonadiabatic}
\end{equation}
is violated are excited nonadiabatically. Fermions cannot occupy a mode with $n_k>1/2$, so unlike the boson case this excitation saturates rather than growing without bound.

This process is governed by a resonance parameter analogous to the one used for bosonic preheating \cite{Kofman:1997yn},
\begin{equation}
q \equiv \frac{\lambda_{\rm Y}^2\Phi^2}{m_\phi^2} \, .
\label{eq:q_def}
\end{equation}
Condition~\eqref{eq:nonadiabatic} fails near $\chi=0$, where $\Omega_k$ changes fastest, so this is where fermion modes are excited. From Eq.~\eqref{eq:fermion_mode_eq}, condition~\eqref{eq:nonadiabatic} requires $\lambda_{\rm Y}^2|\chi||\dot\chi|\gtrsim(k^2+\lambda_{\rm Y}^2\chi^2)^{3/2}$. With $\chi(t)\approx\Phi\cos(m_\phi t)$, $|\dot\chi|\approx m_\phi\Phi$ near a zero-crossing, so, writing $x\equiv|\chi|$, this becomes $k^2\lesssim(\lambda_{\rm Y}^2xm_\phi\Phi)^{2/3}-\lambda_{\rm Y}^2x^2$. Maximizing the right-hand side over $x$ gives the largest momentum for which some $x$ satisfies this condition, $k_{\max}\sim\sqrt{\lambda_{\rm Y} m_\phi\Phi}$. Since excitation occurs for all $k\lesssim k_{\max}$, this sets the width of the excited band, $\Delta k\sim k_{\max}\sim\sqrt{\lambda_{\rm Y} m_\phi\Phi}$, and using $q\equiv\lambda_{\rm Y}^2\Phi^2/m_\phi^2$,
\begin{equation}
\Delta k \sim q^{1/4} m_\phi \, .
\label{eq:band_width}
\end{equation}

Ordinary perturbative decay is recovered only once $q\lesssim1$. For
a real scalar of mass $m_\phi$ coupled as in Eq.~\eqref{eq:yukawa_int},
the tree level decay rate into a fermion pair is
\begin{equation}
\Gamma_{\phi\to\bar\psi\psi} = \frac{\lambda_{\rm Y}^2 m_\phi}{8\pi}\left(1-\frac{4m_\psi^2}{m_\phi^2}\right)^{3/2},
\label{eq:decay_rate_full}
\end{equation}
which in the kinematic limit, $|m_\psi|\ll m_\phi$, becomes,
\begin{equation}
\Gamma_{\phi\to\bar\psi\psi} \simeq \frac{\lambda_{\rm Y}^2 m_\phi}{8\pi}.
\label{eq:decay_rate}
\end{equation}
Because $\Gamma_{\phi\to\bar\psi\psi}$ is fixed by Eq.~\eqref{eq:decay_rate}
while the Hubble rate $H$ falls monotonically as the Universe expands,
the two rates are equal at a unique, finite epoch. In the sudden decay
approximation used throughout this analysis, that epoch defines the
completion of reheating, $\Gamma_{\phi\to\bar\psi\psi} = H(T_{\rm rh})$.
From Eq.~\eqref{eq:Trh}, using $3H_{\rm rh}^2\Mpl^2\approx \rho_{\rm rh}$ we get $H(T_{\rm rh}) = \sqrt{\pi^2 g_{\rm rh}/90}\,T_{\rm rh}^2/M_{\rm Pl}$. This
matching condition fixes the coupling required to reach a given $T_{\rm rh}$,
\begin{equation}
\lambda_{\rm Y}^2 = \frac{8\pi^2}{m_\phi}\sqrt{\frac{g_{\rm rh}}{90}}\,\frac{T_{\rm rh}^2}{M_{\rm Pl}}.
\label{eq:lambda_of_Trh}
\end{equation}
For $N_\star = 55$, $g_{\rm rh}=100$, $\beta=25$ and
$m_\phi\approx 9.7\times10^{13}\,{\rm GeV}$, Eq.~\eqref{eq:lambda_of_Trh}
gives
\begin{align}
T_{\rm rh} &=10^{12}\,{\rm GeV} \;\Rightarrow\; \lambda_{\rm Y} \approx 5.9\times10^{-4},
\label{eq:lambda_12}\\
T_{\rm rh} &=10^{13}\,{\rm GeV} \;\Rightarrow\; \lambda_{\rm Y} \approx 5.9\times10^{-3}.
\label{eq:lambda_13}
\end{align}
Both values are comparable in size to the Standard Model muon to tau
range of charged lepton Yukawa couplings ($\lambda_{\rm Y,\mu}\approx6\times10^{-4}$,
$\lambda_{\rm Y,\tau}\approx1.0\times10^{-2}$) \cite{ParticleDataGroup:2022pth},
so reaching these reheating temperatures through the coupling of
Eq.~\eqref{eq:yukawa_int} does not require an exotic or fine-tuned
interaction strength. This coupling range gives
$\Gamma_{\phi\to\bar\psi\psi}\ll m_\phi$, with
$\lambda_{\rm Y}^2/(8\pi)\sim1.4\times10^{-8}$--$1.4\times10^{-6}$. The field
is weakly damped and undergoes many oscillations before decay
completes, so the oscillation timescale $m_\phi^{-1}$ remains far
shorter than the Hubble timescale at $T_{\rm rh}$. The number of coherent oscillations completed in
one decay time, $N_{\rm osc}\sim m_\phi/(2\pi\Gamma)=4/\lambda_{\rm Y}^2$, is
$\sim10^5$--$10^7$ for the coupling range of
Eqs.~\eqref{eq:lambda_12}--\eqref{eq:lambda_13}.

The perturbative threshold corresponds to $q=1$, {\it i.e.} to a coupling $\lambda_{\rm pert}\equiv m_\phi/\Phi$ set by the amplitude of the first oscillations. Taking $\Phi\approx\phi_{\rm end}-\phi_{\rm eq}\approx0.70\,M_{\rm Pl}\approx1.7\times10^{18}\,{\rm GeV}$ ($\alpha=1$, $\beta=25$) gives $\lambda_{\rm pert}\approx5.7\times10^{-5}$. This uses $m_\phi$ as a fixed reference frequency even during the first oscillation, when the amplitude is not small compared to $\phi_{\rm eq}$ and the potential is not yet quadratic. $q$ and the estimates below are accordingly order of magnitude. The couplings of Eqs.~\eqref{eq:lambda_12}--\eqref{eq:lambda_13} exceed $\lambda_{\rm pert}$ by one to two orders of magnitude, so $q\approx10^2$--$10^4$ at the start of oscillation. The first few oscillations excite fermions through the parametric mechanism of Ref.~\cite{Greene:1998nh}, not directly through Eq.~\eqref{eq:decay_rate_full}.

This early episode removes little energy from $\chi$. Saturating the Pauli bound over the excited band bounds the number density of produced fermions. Following Ref.~\cite{Greene:1998nh}, the comoving number density of fermions in both spin states, including particles and antiparticles, is $n_\psi = \frac{4}{(2\pi)^3}\int d^3k\, n_k$, with $n_k\le1/2$. Approximating the excited region as a sphere of radius $\Delta k$ and saturating $n_k\to1/2$ throughout it,
\begin{equation}
n_\psi \lesssim \frac{4}{(2\pi)^3}\times\frac{4\pi}{3}(\Delta k)^3\times\frac12 = \frac{(\Delta k)^3}{3\pi^2} \, ,
\end{equation}
and using $\Delta k\sim q^{1/4}m_\phi$ from Eq.~\eqref{eq:band_width},
\begin{equation}
n_\psi \lesssim \frac{(\Delta k)^3}{3\pi^2} = \frac{q^{3/4}m_\phi^3}{3\pi^2} \, ,
\end{equation}
which we compare with the number density of $\chi$-quanta in the oscillating field, $n_\phi=\rho_\phi/m_\phi=\tfrac12 m_\phi\Phi^2$.\footnote{For $\chi(t)=\Phi\cos(m_\phi t)$, the energy density is $\rho_\phi=\tfrac12\dot\chi^2+\tfrac12 m_\phi^2\chi^2$. Averaging over one oscillation, $\langle\dot\chi^2\rangle=m_\phi^2\Phi^2\langle\sin^2(m_\phi t)\rangle=\tfrac12 m_\phi^2\Phi^2$ and $\langle\chi^2\rangle=\Phi^2\langle\cos^2(m_\phi t)\rangle=\tfrac12\Phi^2$, so $\rho_\phi=\tfrac14 m_\phi^2\Phi^2+\tfrac14 m_\phi^2\Phi^2=\tfrac12 m_\phi^2\Phi^2$. Since each $\chi$-quantum carries energy $m_\phi$, $n_\phi=\rho_\phi/m_\phi=\tfrac12 m_\phi\Phi^2$.} Using Eq.~\eqref{eq:q_def},
\begin{equation}
\frac{n_\psi}{n_\phi} \lesssim  \frac{2}{3\pi^2}\,q^{3/4}\,\frac{m_\phi^2}{\Phi^2} = \frac{2}{3\pi^2}\,\lambda_{\rm Y}^2\,q^{-1/4} \, ,
\label{eq:fermion_number_bound}
\end{equation}
which for $\lambda_{\rm Y}\approx5.9\times10^{-4}$--$5.9\times10^{-3}$ (corresponding to $q\approx10^2$--$10^4$) gives $n_\psi/n_\phi\lesssim7.43\times10^{-9}$--$2.35\times10^{-7}$. Therefore, only a tiny fraction of the field's quanta converts to fermions during this phase.

Eq.~\eqref{eq:fermion_number_bound} bounds a number density, whereas backreaction depends on the energy density. A produced fermion with $k\lesssim\Delta k$ has energy $\Omega_k=\sqrt{k^2+m_\psi^2(t)}$ with $m_\psi(t)=\lambda_{\rm Y}\chi(t)$, which reaches at most $\lambda_{\rm Y}\Phi=q^{1/2}m_\phi$ as $\chi$ moves away from zero. We use $\Omega_k\lesssim\lambda_{\rm Y}\Phi$ as a conservative upper bound on the energy per fermion. With $\rho_\phi=n_\phi m_\phi$ and Eq.~\eqref{eq:fermion_number_bound},
\begin{equation}
\frac{\rho_\psi}{\rho_\phi}\lesssim\frac{n_\psi}{n_\phi}\,\frac{\lambda_{\rm Y}\Phi}{m_\phi}=\frac{2}{3\pi^2}\,\lambda_{\rm Y}^2\,q^{1/4}\, ,
\label{eq:fermion_energy_bound}
\end{equation}
which gives $\rho_\psi/\rho_\phi\lesssim7.5\times10^{-8}$--$2.4\times10^{-5}$ for $\lambda_{\rm Y}\approx5.9\times10^{-4}$--$5.9\times10^{-3}$ and $q\approx10^2$--$10^4$.

Successive zero crossings do not change this conclusion. Each crossing excites only modes with $k\lesssim\Delta k$, and Pauli blocking limits their occupation to $n_k\le1/2$, so repeated crossings cannot raise the occupation of modes that are already filled. As $q$ decreases with the expansion the excited band narrows, and production stops at $q\simeq1$, Eq.~\eqref{eq:deltaN}.

The excited band, $\Delta k\sim q^{1/4}m_\phi$, starts out at $\Delta k\sim(3\text{--}10)\,m_\phi$ for $q\approx10^2$--$10^4$, above the momentum $k\sim m_\phi/2$ populated by the perturbative channel of Eq.~\eqref{eq:decay_rate}. Since $q\propto\Phi^2\propto a^{-3}$ once $\langle w_\phi\rangle\to0$, and $q$ starts at $q_{\rm i}=(\lambda_{\rm Y}/\lambda_{\rm pert})^2$, the condition $q(N)=q_{\rm i}\,e^{-3N}=1$ gives
\begin{equation}
\Delta N=\tfrac23\ln(\lambda_{\rm Y}/\lambda_{\rm pert}) \, ,
\label{eq:deltaN}
\end{equation}
which is $1.6$--$3.1$ e-folds for the coupling range above. At $N=\Delta N$, $\Delta k\to m_\phi$ (Eq.~\eqref{eq:band_width}). Beyond this point parametric production shuts off and the already produced modes simply redshift kinematically, $k(N)=m_\phi\,e^{-(N-\Delta N)}$. Setting $k=m_\phi/2$ gives $N-\Delta N=\ln2\approx0.69$, so $\Delta k$ falls to $m_\phi/2$ after a further $\ln2$ e-folds, i.e.\ by $N\approx\Delta N+\ln2\approx2.3$--$3.8$, well before the remaining $N_{\rm rh}-\Delta N\sim$ tens of e-folds to $T_{\rm rh}$. The energy removed from the field during this episode scales with the same fraction $n_\psi/n_\phi$, so its backreaction on the equation of motion of $\chi$ is negligible at this level.

Once the early parametric episode has ended, the field decays perturbatively until $H\simeq\Gamma_{\phi\to\bar\psi\psi}$, which marks the completion of reheating in the sudden-decay approximation. Eq.~\eqref{eq:decay_rate} holds for $q\ll1$ and $|m_\psi|\ll m_\phi$, so we check that both conditions are satisfied at this epoch. There $\rho_\phi\simeq3\Gamma^2M_{\rm Pl}^2$, which gives an oscillation amplitude $\Phi^2=2\rho_\phi/m_\phi^2\simeq6\Gamma ^2M_{\rm Pl}^2/m_\phi^2$. Using Eq.~\eqref{eq:decay_rate},
\begin{equation}
q(H\simeq\Gamma)=\frac{\lambda_{\rm Y}^2\Phi^2}{m_\phi^2}\simeq\frac{3\lambda_{\rm Y}^6}{32\pi^2}\,\frac{M_{\rm Pl}^2}{m_\phi^2}\, ,
\label{eq:q_at_decay}
\end{equation}
which is $\approx2.5\times10^{-13}$ for $\lambda_{\rm Y}\approx5.9\times10^{-4}$ and $\approx2.5\times10^{-7}$ for $\lambda_{\rm Y}\approx5.9\times10^{-3}$ (for $M_{\rm Pl}=2.4\times10^{18}\,{\rm GeV}$ and $m_\phi\approx9.7\times10^{13}\,{\rm GeV}$). Hence $q\ll1$ and $|m_\psi|\lesssim\lambda_{\rm Y}\Phi=\sqrt q\,m_\phi\lesssim5\times10^{-4}\,m_\phi$, so the oscillation amplitude has decreased enough by $H\simeq\Gamma$ for the perturbative regime and the limit $|m_\psi|\ll m_\phi$ of Eq.~\eqref{eq:decay_rate} to apply near completion of the decay. The reheating temperatures quoted here assume efficient thermalization of the produced fermions, as in Sec.~\ref{sec:Trh-bookkeeping}, and $H\simeq\Gamma_{\phi\to\bar\psi\psi}$ is a leading-order estimate of the sudden-decay epoch.

The parametric and perturbative regimes are therefore well separated in time. The brief early episode barely dents the energy budget, and the coupling $\lambda_{\rm Y}(T_{\rm rh})$ fixed above from the perturbative matching condition, Eq.~\eqref{eq:lambda_of_Trh}, is the same coupling that governs the much longer perturbative phase, $N_{\rm rh}\gg\Delta N$, that actually completes reheating. The perturbative decay rate and the sudden-decay approximation used in this section therefore describe the bulk of the reheating history.

\subsection{Radiative stability of the inflationary potential}
\label{sec:radstab}

Integrating out $\psi$ at one loop generates a Coleman--Weinberg correction \cite{Coleman:1973jx} to the tree-level potential and its first two derivatives. With the field dependent mass of Eq.~\eqref{eq:mpsi_def}, $m_\psi=\lambda_{\rm Y}\chi$, where $\chi\equiv\phi-\phi_{\rm eq}$, $g_\psi=4$ counts the Dirac fermion's internal degrees of freedom, $\mu$ is the renormalization scale and $L\equiv\ln(\lambda_{\rm Y}^2\chi^2/\mu^2)$, these are
\begin{align}
\Delta V(\phi) &= -\frac{g_\psi}{64\pi^2}\,m_\psi^4\left[\ln\!\left(\frac{m_\psi^2}{\mu^2}\right)-\frac32\right]
\nn \\ &= -\frac{\lambda_{\rm Y}^4\chi^4}{16\pi^2}\left(L-\frac32\right),
\label{eq:CW}\\
\Delta V'(\phi) &= -\frac{\lambda_{\rm Y}^4\chi^3}{4\pi^2}\,\left(L-1\right),
\label{eq:CWprime}\\
\Delta V''(\phi) &= -\frac{\lambda_{\rm Y}^4\chi^2}{4\pi^2}\,\left(3L-1\right).
\label{eq:CWprime2}
\end{align}
These expressions are exact at one loop for the mass function of Eq.~\eqref{eq:mpsi_def}. No assumption $\phi\gg\phi_{\rm eq}$ is made, so they hold over the whole range from $\phi_{\rm end}$ to $\phi_\star$. We fix $\mu=m_\psi(\phi_\star)=\lambda_{\rm Y}\chi_\star$, with $\chi_\star\equiv\phi_\star-\phi_{\rm eq}$, so that $L=0$ at $\phi=\phi_\star$. This gives the potential and its first two derivatives at horizon crossing,
\begin{align}
\Delta V(\phi_\star) &= \frac{3\lambda_{\rm Y}^4\chi_\star^4}{32\pi^2}, \label{eq:CW_at_star}\\
\Delta V'(\phi_\star) &= \frac{\lambda_{\rm Y}^4\chi_\star^3}{4\pi^2}, \label{eq:CWprime_star}\\
\Delta V''(\phi_\star) &= \frac{\lambda_{\rm Y}^4\chi_\star^2}{4\pi^2}. \label{eq:CWprime2_star}
\end{align}

Working within the modified minimal subtraction ($\overline{\text{MS}}$) scheme, the value of $\Delta V(\phi_\star)$ is fixed by the residual scheme-dependent constant. Since we truncate at fixed order without renormalization group improvement or two loop corrections, the result depends on the choice of $\mu$. A different choice reintroduces the logarithm $L$ in Eqs.~\eqref{eq:CW}--\eqref{eq:CWprime2} and changes the correction by a logarithmic factor, without changing its leading scaling with $\lambda_{\rm Y}$ and $\chi_\star$. The bounds derived below should therefore be read as leading order estimates, not exact numbers.

\begin{figure}[t]
\centering
\includegraphics[scale=0.60]{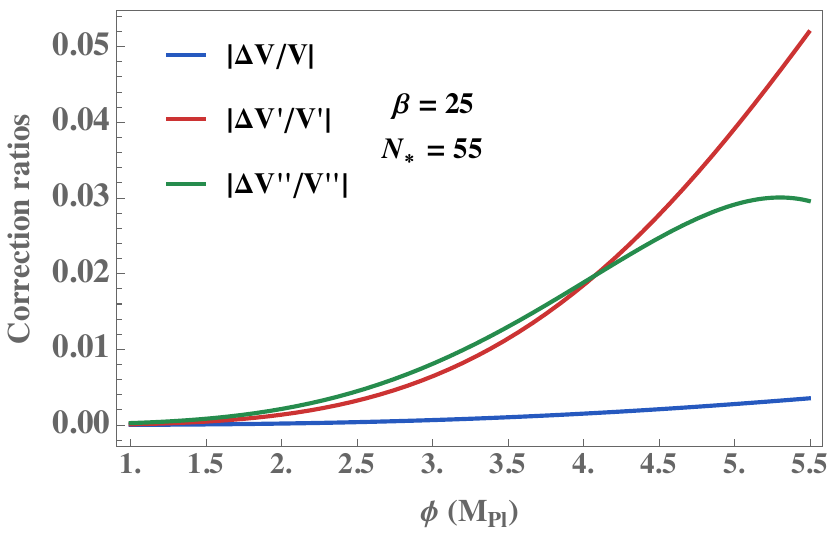}\\
\includegraphics[scale=0.45]{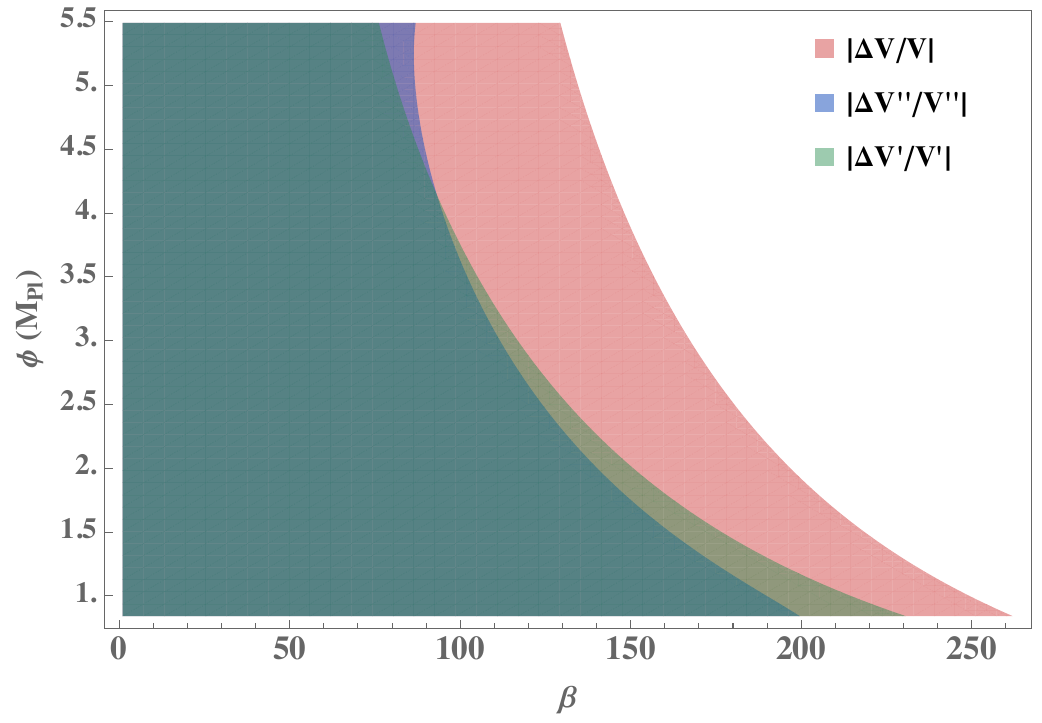}
\caption{\emph{Top:} One-loop correction ratios $|\Delta V/V|$, $|\Delta V'/V'|$, and $|\Delta V''/V''|$, from  Eqs.~\eqref{eq:CW}--\eqref{eq:CWprime2}, are shown for $\beta=25$, $N_\star=55$, $\alpha=1$ and $g_{\rm rh}=100$. The range shown, $\phi_{\rm end}\lesssim\phi\lesssim\phi_\star$, covers the field excursion during inflation itself, from the horizon crossing to end of inflation. \emph{Bottom:} shaded regions in the $(\beta,\phi)$ plane represent the region where each of the ratio from Eqs.~\eqref{eq:CW}--\eqref{eq:CWprime2} $\leq 1$ for $N_\star=55$, $\alpha=1$, $T_{\rm rh}=10^{12}\,{\rm GeV}$  and $g_{\rm rh}=100$.}
\label{fig:radstab_field_beta}
\end{figure}

Equations~\eqref{eq:CW_at_star}--\eqref{eq:CWprime2_star} evaluate the correction only at the single field value $\phi_\star$. Since the slow roll parameters must stay controlled over the full range of $\phi$ values during the last $N_\star=50$--60 e-folds, not just at horizon crossing, Fig.~\ref{fig:radstab_field_beta} evaluates the ratios of Eqs.~\eqref{eq:CW}--\eqref{eq:CWprime2}. In the upper panel, the ratio involving the first derivative of the radiative correction, $\Delta V'/V'$ in Eq.~\eqref{eq:CWprime}, increases more rapidly near horizon crossing than the other ratios. However, all the ratios remain well below unity for $\beta=25$ and $N_\star=55$. The lower panel of Fig.~\ref{fig:radstab_field_beta} shows the allowed region in the $\beta$--$\phi$ plane for which all the ratios in Eqs.~\eqref{eq:CW}--\eqref{eq:CWprime2} remain $\leq 1$ for $T_{\rm rh}=10^{12}\,{\rm GeV}$. This condition places an upper bound on $\beta$, with $\beta\lesssim 76$. Since the post-inflationary dynamics gives a lower bound on $\beta$, the radiative corrections provide an upper bound. For $N_\star=55$, we therefore obtain approximately $0.71\lesssim \beta \lesssim 76$.

As a first criterion for radiative stability, we require $|\Delta V(\phi_\star)|\lesssim$ $V(\phi_\star)$. A correction comparable to or larger than the tree level potential can significantly change the inflationary dynamics. Substituting $\lambda_{\rm Y}^2$ from Eq.~\eqref{eq:lambda_of_Trh} into Eq.~\eqref{eq:CW_at_star} and solving $|\Delta V(\phi_\star)|=V(\phi_\star)$ for $T_{\rm rh}$ gives the crossover temperature above which this criterion fails,
\begin{equation}
T_{\rm rh,c} = \left(\frac{15\, m_\phi^2\, M_{\rm Pl}^2\, V(\phi_\star)}{\pi^2 g_{\rm rh}\,{\chi_\star^4}}\right)^{1/4}.
\label{eq:Trh_crossover}
\end{equation}
The slow-roll parameters, however, depend on $V'$ and $V''$, not just $V$, and a correction can distort these before it is comparable to $V(\phi_\star)$ itself. Fig.~\ref{fig:radstab_field_beta} shows that the radiative correction to the first derivative grows fastest near horizon crossing, so we repeat the calculation for the stricter condition $|\Delta V'(\phi_\star)|=|V'(\phi_\star)|$. Using Eq.~\eqref{eq:CWprime_star} and $V'(\phi_\star)=\alpha M_{\rm Pl}V(\phi_\star)/\phi_\star^2$ we get
\begin{equation}
T_{\rm rh,c}' = \left(\frac{45\, m_\phi^2\, M_{\rm Pl}^3\, \alpha\, V(\phi_\star)}{8\pi^2 g_{\rm rh}\,{\phi_\star^2\chi_\star^3}}\right)^{1/4}.
\label{eq:Trh_crossover_slope}
\end{equation}
$T_{\rm rh,c}$ compares the correction to the potential $V(\phi_\star)$, while $T_{\rm rh,c}'$ compares it to the tree level slope $V'$ at the same field value $\phi_\star$. For $N_\star=55$, $g_{\rm rh}=100$, and $m_\phi=1.3\times10^{14}$, $9.7\times10^{13}$, $8.3\times10^{13}\,{\rm GeV}$ for $\beta=15,\; 25,\; 30$, Eqs.~\eqref{eq:Trh_crossover} and \eqref{eq:Trh_crossover_slope} give
\begin{align}
&T_{\rm rh,c}\approx{4.9,\,4.2,\,3.9}\times10^{12}\,{\rm GeV} \; , 
\label{eq:Trhc_values}\\ 
& T_{\rm rh,c}'\approx{2.5,\,2.1,\,2.0}\times10^{12}\,{\rm GeV} \; ,
\label{eq:Trhc_prime_values}
\end{align}
respectively, with $T_{\rm rh,c}'$ roughly a factor of two below $T_{\rm rh,c}$.

\begin{figure}[t]
\centering
\includegraphics[scale=0.60]{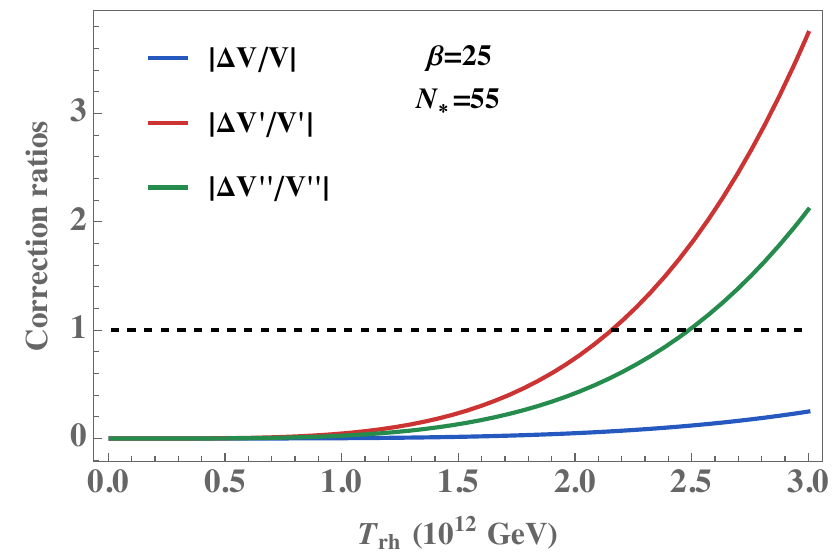}
\caption{Ratios $|\Delta V/V|$, $|\Delta V'/V'|$, and $|\Delta V''/V''|$ of the one-loop Coleman--Weinberg correction, Eq.~\eqref{eq:CW}, to the tree-level potential and its first two derivatives, as functions of the reheating temperature $T_{\rm rh}$, for $N_\star=55$, $\beta=25$ and $\alpha=1$.}
\label{fig:radstab}
\end{figure}

\begin{figure}[t]
\centering
\includegraphics[scale=0.60]{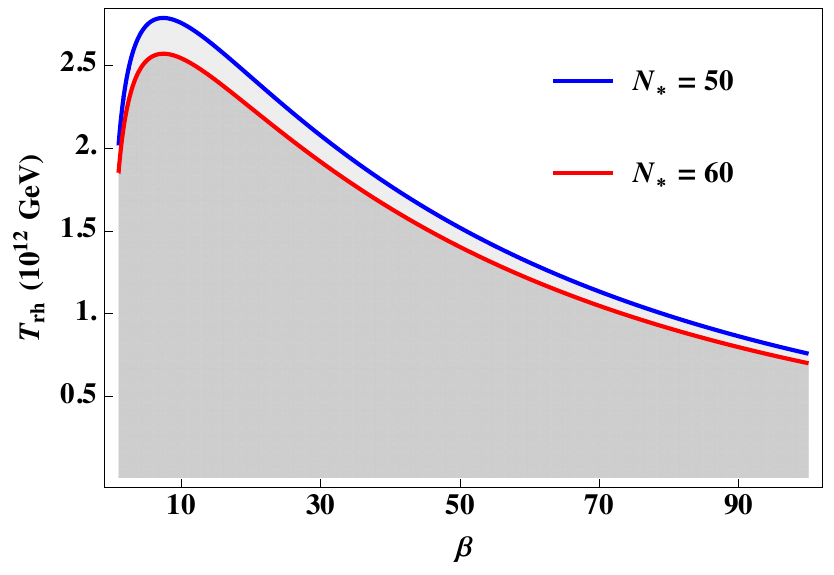}
\caption{Boundary in the $(\beta,T_{\rm rh})$ plane, for $\alpha=1$, below which $|\Delta V'/V'|\le1$, shown for $N_\star=50$ and $N_\star=60$. The curve is $T_{\rm rh,c}'(\beta)$ of Eq.~\eqref{eq:Trh_crossover_slope} and gives an approximate upper bound on the radiatively safe reheating temperature at each $\beta$.}
\label{fig:Trhc-beta}
\end{figure}

Figure~\ref{fig:radstab} shows $|\Delta V/V|$, $|\Delta V'/V'|$, and $|\Delta V''/V''|$ as functions of $T_{\rm rh}$, for $N_\star=55$, $\beta=25$ and $\alpha=1$. $|\Delta V'/V'|$ grows fastest and crosses unity at $T_{\rm rh}\simeq2.1\times10^{12}\,{\rm GeV}$, below $T_{\rm rh,c}$ from Eq.~\eqref{eq:Trh_crossover}. Comparing only $|\Delta V(\phi_\star)|$ with $V(\phi_\star)$ therefore understates how quickly the correction affects the slow roll parameters, and we take $T_{\rm rh,c}'$, not $T_{\rm rh,c}$, as the operative bound.

At $\phi_\star$, the tree level potential gives $V'(\phi_\star)=\alpha M_{\rm Pl}V(\phi_\star)/\phi_\star^2$ exactly, and, from Eq.~\eqref{eq:ep_eta_V}, $V''(\phi_\star)=-\left(2-\alpha M_{\rm Pl}/\phi_\star\right)\alpha M_{\rm Pl}V(\phi_\star)/\phi_\star^3$, where we neglect the subtracted exponential of Eq.~\eqref{eq:Vsquared}, which is negligible at $\phi_\star$. Combined with Eqs.~\eqref{eq:CW_at_star}--\eqref{eq:CWprime2_star}, this gives, independent of $\lambda_{\rm Y}$,
\begin{align}
\frac{|\Delta V'/V'|}{|\Delta V''/V''|} & =\frac{\chi_\star}{\phi_\star}\left(2-\frac{\alpha M_{\rm Pl}}{\phi_\star}\right)\simeq1.8, \\  \frac{|\Delta V'/V'|}{|\Delta V/V|} &=\frac{8}{3}\frac{\phi_\star^2}{\alpha M_{\rm Pl}\chi_\star}\simeq15.
\label{eq:ratio_hierarchy}
\end{align}
For $\alpha=1$ and $N_\star=55$, $\phi_\star\approx5.5\,M_{\rm Pl}$ (Eq.~\eqref{eq:phi_end_star}), giving $|\Delta V'/V'|\approx15\,|\Delta V/V|$, consistent with Fig.~\ref{fig:radstab}. Since $\phi_\star\propto N_\star^{1/3}$ at fixed $\alpha$, the ratio in Eq.~\eqref{eq:ratio_hierarchy} is 14.6 at $N_\star=50$ and 15.5 at $N_\star=60$, so it changes by only about $6\%$ between $N_\star=50$ and $60$. The hierarchy $|\Delta V'/V'|>|\Delta V''/V''|>|\Delta V/V|$ and the choice of $T_{\rm rh,c}'$ therefore holds throughout the $e$-fold range relevant to observations.

Figure~\ref{fig:Trhc-beta} extends the comparison by considering the allowed region in the $\beta-T_{\rm rh}$ plane. For each $\beta$ over the range shown, it shows the boundary $T_{\rm rh,c}'(\beta)$ of Eq.~\eqref{eq:Trh_crossover_slope}, at $\alpha=1$, for $N_\star=50$ and $N_\star=60$. The bound follows $m_\phi$, which peaks at $\beta\approx8$ for $\alpha=1$, so it varies little for $\beta\lesssim10$ and falls for larger $\beta$, and the two $N_\star$ curves stay close together over the full range, confirming that this bound is not sensitive to the choice of horizon crossing epoch.

The condition $|\Delta V'/V'|=1$ defines the temperature $T'_{\rm rh,c}$ at which the loop correction becomes comparable to the tree level slope. It is an order unity boundary and does not by itself imply small corrections below it. Since $\Delta V\propto\lambda_{\rm Y}^4\propto T_{\rm rh}^4$, the ratios at $\phi_\star$ for $T_{\rm rh}=10^{12}\,{\rm GeV}$ ($\alpha=1$, $\beta=25$, $N_\star=55$) are $|\Delta V/V|\simeq0.003$, $|\Delta V'/V'|\simeq0.05$ and $|\Delta V''/V''|\simeq0.03$, and each is larger by a factor $10^4$ at $T_{\rm rh}=10^{13}\,{\rm GeV}$, the upper end of the range obtained in Sec.~\ref{sec:Trh-bookkeeping}. We therefore regard $T_{\rm rh}\lesssim10^{12}\,{\rm GeV}$ as radiatively safe, since all three corrections are then small compared with the tree level terms. As $T_{\rm rh}$ approaches $T'_{\rm rh,c}\simeq2\times10^{12}\,{\rm GeV}$ the one loop corrected slow roll parameters can no longer be trusted at leading order.

\section{Conclusions}
\label{sec:conc}
  
We have studied a single-field inflationary model based on the inverse exponential (IExp) potential of Eq.~\eqref{eq:potential_inflation}, which depends on a single free parameter $\alpha$ besides the overall energy scale $V_0$. For $N_\star=50$--$60$ e-folds and $\alpha$ up to a few, the predicted scalar spectral index and tensor to scalar ratio lie within the $1\sigma$ SPA+BK+DESI2 region (Figs.~\ref{fig:inflation_parameters} and \ref{fig:r-ns}), whose constraints are $n_s=0.9728\pm0.0029$ and $r<0.035$ \cite{Balkenhol:2025wms}. Larger $\alpha$ raises $r$, and for $\alpha\gtrsim7$--$10$ it exceeds this bound. The running of $n_s$ stays negligible over the same range. We related this behaviour to the curvature parameter $\Gamma$ of the potential. Viable inflation requires a small slope $\lambda$ together with moderate to large negative $\Gamma$, that is, a strongly concave potential (Appendix~\ref{app:lambdaGamma}).
 
This agreement with observations is not a special property of the exponent $n=-1$ in the more general family $V\propto e^{-\alpha(\phi/M_{\rm Pl})^n}$, the generalised IExp~\eqref{eq:potential_inflation_general}. Scanning $-10\le n\le-1$ at fixed $\alpha$ shifts $n_s$ by only a few $\times10^{-3}$ while $r$ falls by roughly two orders of magnitude, and almost the entire scanned region stays within $1\sigma$ of SPA+BK+DESI2 (Fig.~\ref{fig:r-ns-genn}). The case $n=-1$ is therefore singled out by its simplicity, a single free parameter beyond $V_0$, and not by any tuning of the $(n_s,r)$ prediction. Compared with the $R^2$ (Starobinsky) model, monomial potentials, $\alpha$-attractor E- and T-models, non-minimally coupled Higgs like inflation and a radiatively-corrected plateau potential (Sec.~\ref{sec:comparison}), the IExp prediction for $\alpha$ up to a few lies inside the $1\sigma$ SPA+BK+DESI2 contour, the $R^2$ prediction lies near the boundary of the $2\sigma$ region, and the plain monomial family is excluded over most of $0.01\le n\le5$. The non-minimally coupled Higgs like and radiatively corrected plateau families also overlap the favoured contours over part of their parameter space.
 
To extend the model beyond inflation, we added a second, steep exponential term to the potential, Eq.~\eqref{eq:Vsquared}, which is negligible during inflation but generates a minimum at $\phi_{\rm eq}=\sqrt{\alpha/(2\beta)}\,M_{\rm Pl}$ once the field rolls past $\phi_{\rm end}$, provided $\beta>1/\sqrt2$. Solving the Klein--Gordon equation numerically from the end of inflation shows the field then oscillates about this minimum and reheats the Universe, with the time averaged equation of state set by $\beta$ (Fig.~\ref{fig:post-inf}). Using standard reheating formalism, we obtained the reheating e-fold number and temperature for several assumed equations of state $w_{\rm rh}$ (Fig.~\ref{fig:reheating}). The resulting $T_{\rm rh}$ remains consistent with the $1\sigma$ SPA+BK+DESI2 bound on $n_s$ across the sampled range of $\alpha$, with typical maximum values $T_{\rm rh}\sim10^{12}$--$10^{13}\,{\rm GeV}$, though the low-temperature tail of this window is excluded once the BBN bound $T_{\rm rh}\gtrsim3.67\,{\rm MeV}$ is imposed.

We then checked whether these reheating temperatures are reachable through an explicit, minimal decay channel, and whether the coupling this requires is compatible with the flatness of the inflationary potential. Coupling the field to a Dirac fermion through a single renormalizable Yukawa interaction, with the fermion mass sourced entirely by the same coupling near the post inflationary minimum and at horizon crossing (Sec.~\ref{sec:yukawa}), reproduces $T_{\rm rh}\sim10^{12}$--$10^{13}\,{\rm GeV}$ for couplings $\lambda\sim5.9\times10^{-4}$--$5.9\times10^{-3}$, comparable in size to Standard Model charged lepton Yukawa couplings, with the sudden-decay and perturbative width approximations shown, at the order of magnitude level, to hold over this range (Sec.~\ref{sec:yukawa}). The same coupling, however, generates a one loop Coleman--Weinberg correction to the potential during inflation (Sec.~\ref{sec:radstab}). Comparing the correction not only to $V$ but to $V'$ and $V''$ across the full $N_\star=50$--60 field range shows that $\Delta V'/V'$ grows fastest near horizon crossing (Figs.~\ref{fig:radstab_field_beta} and \ref{fig:radstab}), so it is this stricter ratio, rather than $\Delta V/V$ alone, that sets the operative bound. Requiring all three ratios to remain below unity over this range restricts the potential parameter $\beta$ to $1/\sqrt{2}\lesssim\beta\lesssim 76$ for $T_{\rm rh}=10^{12}\,{\rm GeV}$, the lower limit set by the graceful exit condition of Sec.~IV and the upper limit by radiative stability. Within this window, the $\Delta V'/V'$ bound gives a crossover temperature $T_{\rm rh,c}'\approx{2.5,\,2.1,\,2.0}\times10^{12}\,{\rm GeV}$ for $\beta=15,\,25,\,30$ respectively, so that $T_{\rm rh}\lesssim10^{12}\,{\rm GeV}$ keeps all three one loop ratios at or below about 0.05 at horizon crossing, while temperatures approaching $10^{13}\,{\rm GeV}$ give ratios far above unity.
 
Overall, the IExp potential gives a minimal, single parameter description of inflation consistent with the SPA+BK+DESI2 constraints. Its extension by a second exponential term gives a calculable reheating phase with temperatures that respect the BBN bound, while radiative stability restricts the Yukawa mediated reheating temperature to $T_{\rm rh}\lesssim10^{12}\,{\rm GeV}$. The extra parameter $\beta$ is introduced to create the minimum and is not fixed by the inflationary dynamics, so the reheating results belong to the extended model and are not predictions of the IExp potential alone.

\appendix

\section{Slope and curvature parameters of the inflationary potential}
\label{app:lambdaGamma}

The slope and curvature parameters of the potential are defined as
\begin{eqnarray}
\lambda = -\frac{M_{\rm Pl} V'}{V}, \qquad
\Gamma = \frac{V''V}{V'^2}\, .
\end{eqnarray}
In terms of these parameters, the slow-roll quantities become $\epsilon_V = \lambda^2/2$ and $\eta_V = \lambda^2 \Gamma$.
The tensor-to-scalar ratio and scalar spectral index can then be written as
\begin{eqnarray}
n_s = 1 - \lambda^2(3 - 2\Gamma), \qquad
r = 8\lambda^2.
\end{eqnarray}

\begin{table}[ht]
\centering
\caption{Allowed values of the potential flow parameters $\lambda$ and $\Gamma$, together with the corresponding slow-roll parameters $\epsilon_V$ and $\eta_V$, for representative values of $r$ assuming $0.95<n_{\rm s}<0.98$.}
\label{tab:lambda-gamma-values}
\resizebox{\columnwidth}{!}{%
\begin{tabular}{c|c|c|c}
\hline
 & \multirow{2}{*}{$r = 0.034$} &
$\epsilon_V \simeq 0.0021$ & $\lambda \simeq 0.065$ \\
 & & $-0.018 \lesssim \eta_V \lesssim -0.0038$ &
$-4.4 \lesssim \Gamma \lesssim -0.9$\\
\cline{2-4}
\multirow{2}{*}{$0.95<n_{\rm s}<0.98$} &
\multirow{2}{*}{$r = 0.01$} &
$\epsilon_V \simeq 0.00063$ & $\lambda \simeq 0.035$ \\
 & & $ -0.023 \lesssim \eta_V \lesssim -0.0081$ &
$-18.5 \lesssim \Gamma \lesssim -6.5$\\
\cline{2-4}
 & \multirow{2}{*}{$r = 0.005$} &
$\epsilon_V \simeq 0.00031$ & $\lambda \simeq 0.025$ \\
 & & $ -0.024 \lesssim \eta_V \lesssim -0.0091$ &
$-38.5 \lesssim \Gamma \lesssim -14.5$\\
\hline
\end{tabular}}
\end{table}

Using observational constraints on $n_s$ and representative values of $r$, one can map the allowed regions in $(\lambda,\Gamma)$ space. The resulting ranges are summarized in Table~\ref{tab:lambda-gamma-values}. These results indicate that viable single field inflation requires a small slope parameter together with moderate to large negative $\Gamma$, implying that the inflationary potential must be strongly concave.

\section{Reheating e-folds}
\label{app:reheat}

At the end of inflation, when the equation-of-state (EoS) approaches $w = -1/3$, the total energy density is related to the potential energy as $\rho_{\rm end} = \tfrac{3}{2} V_{\rm end}$. The number of e-folds during reheating, $N_{\rm rh}$, defined as the interval between the end of inflation and the onset of the radiation-dominated epoch ($w = 1/3$), is then
\begin{align}
N_{\rm rh} &= \ln\!\left(\frac{a_{\rm rh}}{a_{\rm end}}\right)
= \frac{1}{3(1+w_{\rm rh})}
\ln\!\left(\frac{\rho_{\rm end}}{\rho_{\rm rh}}\right) \nonumber \\
&= \frac{1}{3(1+w_{\rm rh})}
\ln\!\left(\frac{3}{2}\frac{V_{\rm end}}{\rho_{\rm rh}}\right).
\label{eq:Nrh_def}
\end{align}

For $g_{\rm rh}\simeq 100$,
\begin{equation}
\ln \left[ \left(\frac{43}{11 g_{\rm rh}}\right)^{1/3} \frac{a_0 T_0}{k_\star} \right] \approx 61.6 \; ,
\label{eq:61.6}
\end{equation}
where $a_0$ and $T_0 = 2.73~\mathrm{K}$ denote the present-day scale factor and CMB temperature, and $k_\star = 0.05~\mathrm{Mpc}^{-1}$ is the pivot scale. Using this result, the reheating e-fold number can be expressed as in Eq.~\eqref{eq:Nrh}.

\bibliographystyle{apsrev4-2} 
\bibliography{references}

\end{document}